\documentclass[final,3p,12pt,times]{elsarticle}
\usepackage{color}
\usepackage{graphicx}
\usepackage{amsmath, mathtools}
\usepackage{amssymb}
\usepackage{graphicx}
\usepackage{grffile}
\usepackage{epstopdf}
\usepackage{enumitem} 
\usepackage{setspace}
\usepackage{subfigure}
\usepackage{natbib}
\usepackage{lscape}
\usepackage{soul}
\usepackage{booktabs}
\graphicspath{{./Figs/}}
\usepackage{upgreek}
\newcommand{\rb}{\textcolor{black}} 
\usepackage{gensymb} 
 \usepackage{nomencl} 
\makenomenclature

\usepackage{stackengine}
\setstackgap{S}{2pt}
\setstackgap{L}{.7\baselineskip}

\journal{Springer Nature Applied Sciences}
\usepackage{amssymb}

\begin{document}

\begin{frontmatter}
\title{Computational Modeling and Analysis of Flow-induced Vibration of an Elastic Splitter Plate Using a Sharp-interface Immersed Boundary Method}
\author[1]{Anup Kundu}
\author[2]{Atul K. Soti}
\author[3]{Hemanshul Garg}
\author[3]{Rajneesh Bhardwaj\corref{mycorrespondingauthor}}
\cortext[mycorrespondingauthor]{Corresponding author:}
\ead{rajneesh.bhardwaj@iitb.ac.in}
\author[4]{Mark C. Thompson}

\address[1]{C. V. Raman College of Engineering, Bhubaneswar 752054, India}
\address[2]{Department of Mechanical Engineering, Indian Institute of Guwahati, Guwahati 781039 India}
\address[3]{Department of Mechanical Engineering, Indian Institute of Technology Bombay, Mumbai 400076, India}
\address[4]{Department of Mechanical and Aerospace Engineering, Monash University, Melbourne 3800, Australia.}

\begin{abstract}
We present the development and benchmarking of an in-house fluid-structure interaction (FSI) solver. An implicit partitioned approach is utilized to couple a sharp-interface immersed boundary (IB) method based flow solver and a finite-element method based structural solver. In the present work, the coupling is accelerated using a dynamic under-relaxation scheme. The revised coupling is around two to three times faster and numerically stable, as compared to the one that uses a constant under-relaxation parameter. The solver is validated against two FSI benchmarks in which a thin, finite thickness, elastic splitter plate is attached to the lee side of a circular or square rigid cylinder, subjected to laminar flow. In these two-dimensional benchmarks, the flow induces a wave-like deformation in the plate, and it attains a periodic self-sustained oscillation. We employ the FSI solver to analyze the flow-induced vibration (FIV) of the plate in a uniform laminar free-stream flow for a wide range of mass ratio and bending stiffness at Reynolds number ($Re$) of 100, based on the diameter of the cylinder. At the given $Re$, two-dimensional numerical simulations show that the FIV of the plate effectively depends only on the mass ratio and bending stiffness. The largest displacement of the plate vibration is found to occur in the lock-in region, where the vortex shedding frequency of the coupled fluid-structure system is close to the natural frequency of the splitter plate. We briefly discuss wake structures and phase plots for different cases of mass ratio and bending stiffness. 
\end{abstract}
\end{frontmatter}

\section{Introduction}\label{intro}

Flow-induced vibration (FIV) of an elastic plate subjected to laminar flow has potential applications in energy harvesting \cite{BHATTACHARYA2016140, asoti2017} and thermal augmentation \cite{shoele2014computational, soti2015flow, joshi2015numerical}. The interaction of fluid flow with a flexible structure may lead to large-scale FIV due to the resonant forcing of the structure caused by periodic vortex shedding. Fluid-structure interaction (FSI) modeling of large-scale FIV poses a significant challenge of tackling a deforming structure in a fluid domain. Moreover, geometric and/or material non-linearity should be accounted for in the structural solver. The non-linear system of governing equations of the fluid and structure may be strongly coupled to accurately capture large-scale FIV. Such non-linear FSI systems exhibit a large amplitude of FIV over a wide range of flow velocity \cite{BADHURSHAH2019102697}, potentially useful in broadband energy harvesting devices. 

Previous studies attempted the computational modeling of moving structure in a fluid domain using either immersed boundary (IB) method or arbitrary Lagrangian-Eulerian (ALE) method. In the latter, a structure-conformal grid is used and it gets distorted at each time step due to flow-induced deformation (FID) of the structure. Therefore, a new mesh should be generated and the numerical solution should be mapped to this new grid. By contrast, the former is well-suited to address this computational challenge as compared to the latter. A structure non-conformal (oftentimes a Cartesian) grid is used in the IB method and there is no need to remesh the fluid domain while tackling a moving structure boundary. A review of variants of IB methods is provided by Mittal and Iaccarino \cite{mittal2005immersed} \rb{ and Sotiropoulos and Yang \cite{SOTIROPOULOS20141}.}

In order to account structural dynamics in an FSI system, previous studies successfully integrated finite-element based structural solver with existing flow solvers. For example, Bhardwaj and Mittal \cite{bhardwaj2012benchmarking} proposed an FSI solver by coupling a sharp-interface IB method and an open-source finite-element solver (Tahoe), using an implicit partitioned approach. Employing this solver, they validated the FSI benchmark, proposed by Turek and Hron \cite{turek2006proposal}. In this benchmark, an elastic plate attached to a rigid cylinder attains self-sustained oscillation in a laminar channel flow. Similarly, Tian et al. \cite{TIAN2014451} proposed a versatile FSI solver which could handle large-scale FID of a flexible structure. They carried out several validations with established benchmarks and demonstrated the three-dimensional capability of the solver. Bailoor et al. \cite{bailoor2017fluid} coupled a compressible flow solver with an open-source finite-element solver (Tahoe) to simulate blast loading on thin plates. Very recently, Furquan and Mittal \cite{furquan2015flow} numerically studied two side-by-side flexible splitter plates attached to square cylinders using a Deforming-Spatial-Domain/Stabilized Space-Time flow solver coupled with a finite-element open-source structural dynamics solver. 

\rb{The dynamic relaxation using Aitken's method was employed to accelerate the convergence of the coupling between the flow and structural solver in previous reports. K\"{u}ttler and Wall \cite{wall2008} demonstrated a successful implementation of the dynamic relaxation using Aitken's method in an FSI solver based on the ALE method. They showed a reduction in the number of sub-iterations by two to four times as compared to the constant under-relaxation scheme. Similarly, Borazjani et al. \cite{BORAZJANI2} implemented Aitken's method in an IB method based flow solver coupled with an elastically mounted rigid structure. Later, Kim et al. \cite{KIM2018296} reported the implementation of the Aitken's method for an IB method based flow solver and a structural dynamics solver, coupled using weak and strong couplings. Degroote et al. \cite{degroote2010performance} also presented a detailed algorithm that employed Aitken’s method.} 

\rb{Several previous reports elucidated the flow physics of a {\textit{rigid}} splitter plate mounted on a cylinder. Vu et al. \cite{Vu2016} numerically examined the effects of splitter plate length and Reynolds numbers on flow characteristics and drag/lift coefficients at $Re = 60-180$. They found that a critical plate length exists to suppress the vortex shedding. Sarioglu \cite{SARIOGLU2017221} measured flow-field around a rigid splitter plate mounted on a square cylinder at $Re = 30,000$, keeping the plate length equal to cylinder diameter. The author varied the angle of incidence and reported a large Strouhal number and the lowest drag at an angle of 13$^{\circ}$. Similarly, Chauhan et al. \cite{CHAUHAN2018319} experimentally measured the flow field around a rigid plate mounted on a square cylinder at  $Re = 485$. They varied plate length from 0 to 6 times of the cylinder width and reported that a secondary vortex appears near the tailing edge of the plate for a threshold plate length. } 

Several previous studies \cite{connell2007flapping, alben2015flag, akcabay2012hydroelastic, liu2014stable} defined two important dimensionless parameters that govern the dynamics of a {\textit{flexible}} plate subjected to fluid flow, namely, the bending stiffness ($K_{b}$) and the mass ratio ($M$). These are given by 
\begin{equation} \label{eq:1}
 {K_{b}=\frac{E^*}{\rho_{f}^{*}{{{U^*_{\infty}}}^2}}\frac{{h^*}^3}{12L^*{^3}}},
\end{equation} 
\begin{equation} \label{eq:2}
M=\frac{\rho_{p}^{*} h^*}{\rho_{f}^{*} L^*},
\end{equation}
where the superscript $^*$ denotes a dimensional variable. Here, $E^*$, $\rho_{f}^*$, ${U^*_{\infty}}$, $h^*$, $L^*$, $\rho_{p}^*$ are the Young\textsc{\char13}s modulus of the plate, fluid density, free-stream velocity, plate thickness, plate length and plate density, respectively. Note that $K_b$ is defined per unit spanwise width of the plate in eq.~\ref{eq:1}. Physically, $K_{b}$ represents the ratio of restoring force produced by stiffness to the loading on the structure by the fluid. The parameter $M$ represents the ratio of the density of the structure to that of the fluid, which is often referred to as the mass ratio. In addition, the reduced velocity $U_R$ is another important parameter, defined as the ratio of the characteristic time scale of the structure to that of the fluid \cite{tang2007instability} and is given by,
\begin{equation} \label{eq:3}
U_R=\sqrt{\frac{ M}{{K_{b}}}}.
\end{equation}

The dynamics of thin, flexible plate oscillations subjected to a free-stream flow has been reported in several studies. Watanabe et al. \cite{watanabe2002theoretical} studied the flutter of a paper sheet using an analytical method and reported high flutter modes at low $M$. Argentina and Mahadevan \cite{argentina2005fluid} proposed a critical speed for the onset of flapping and estimated the flapping frequency based on scaling analysis. Tang and Pa\"idoussis \cite{tang2007instability} investigated the dynamics of a flexible plate using a Euler-Bernoulli model coupled with an unsteady lumped-vortex model. They investigated the flutter boundary and the post-critical behavior of this FSI system. They obtained the flutter boundary in the form of the critical flow velocity versus the length of the flexible plate. They observed that the critical flow velocity is sensitive to short plate lengths. 

Previous studies reported different regimes of flapping or flutter of a thin plate. A comprehensive review of such regimes was provided by Shelley and Zhang \cite{shelley2011flapping}. Connell and Yue \cite{connell2007flapping} proposed a regime map of flag flutter based on their FSI simulations. They proposed the following three categories of plate dynamics: fixed-point stability, limit-cycle flapping, and chaotic flapping. Fixed-point stability occurs when the flag aligns with the flow. As the flow velocity is increased, limit-cycle flapping takes over, characterized by single-frequency repeating flag oscillations. Chaotic flapping occurs as the flow velocity is further increased. Similarly, Lee et al. \cite{lee2014flapping} examined the flapping dynamics of a flexible flag in a uniform flow. They found three different flapping states such as regular flapping, irregular flapping and irregular flapping with violent snapping by varying $M$ and $K_b$. 

Alben and Shelley \cite{alben2008flapping} simulated the nonlinear dynamics of a flexible sheet in a 2D inviscid fluid. They characterized the behavior of flapping flags at large amplitudes and over many flapping periods and demonstrated a transition from a periodic to a chaotic flapping as the bending rigidity was decreased. They also found that the stability boundary of the flow-aligned state for a flag within the two-dimensional parameter space of dimensionless flag inertia and bending rigidity. Employing a linear stability analysis, Connell and Yue \cite {connell2007flapping} found the existence of a critical mass ratio for the chaotic flapping of the plate. Similarly, Eloy et al. \cite{eloy2007flutter} studied the linear stability of a flexible plate immersed in axial flow and found that a finite-span plate is stable than the infinite-span plate. Eloy et al. \cite{eloy2008aeroelastic} addressed the linear stability of the rectangular plate in uniform flow and incompressible axial flow by varying aspect ratio. They identified critical velocities for the instability transitions as a function of system parameters, showing good agreement with their data. 

Akcabay and Young \cite{akcabay2012hydroelastic} examined the dynamic response and stability of piezoelectric beams in viscous and axial flows. They showed that a heavy beam undergoes flutter in a light fluid when the fluid inertial forces are in the balance with the solid elastic restoring forces, and for a light beam in a heavy fluid, flutter occurs when the fluid inertial force dominates the solid inertial force. Recently, Shoele and Mittal \cite{shoele2014computational} numerically studied the dynamics of a self-oscillating reed in a channel flow, and they found that heavy reeds have higher critical velocities, and have low oscillations frequencies and amplitudes. In a follow-up study, they predicted the flutter instability inside for a plate confined in a 2D channel of height $H$ on the $M$-$U_R$ plane for a channel length $L = 1$, for $Re = 400$ \cite{shoele2016flutter}. They found that confinement induces a destabilizing effect and increases the oscillation frequency and compared the stability curves for different values of the $H/L$ ratio. Their study found that using confinement and the asymmetric position of the plate could be used to adjust the flutter frequency and flutter instability. 

In the context of an elastic splitter plate mounted on a rigid cylinder, \rb{Turek and Hron \cite{turek2006proposal} proposed a FSI benchmark in which the elastic splitter plate of aspect ratio 17.5 attains a self-sustained periodic oscillation in a channel flow. Using the same configuration of Turek and Hron \cite{turek2006proposal}, Bhardwaj and Mittal \cite{bhardwaj2012benchmarking} numerically showed that the oscillation frequency of the plate varies linearly with the dilatational wave speed inside the plate. Kundu et al. \cite{kundu2017response} showed that the computed plate frequency in the lock-in regime scales as the second mode of the natural frequency of a vibrating cantilevered plate ($\mathit{f_{ni}^*}$) in the vacuum. In this context,} the natural frequency of an elastic plate fixed at one end is obtained using the Euler-Bernoulli beam model and is given by \cite{ kundu2017response,thomson1996theory},
\begin{equation} \label{eq:4}
\mathit{f_{ni}^*}=\frac{k_{i}^2}{2\pi}\sqrt{\frac{E^{*}I^{*}}{\rho_{s}^*A^*{L^*}^4}},
\end{equation}
where $i$ = 1, 2, 3 represents the frequency modes of the plate, $E^{*}I^{*}$ is the dimensional flexural rigidity of the beam and $k_{i}$ are the respective constants for the modes. The values of $ k$ are 1.875, 4.694 and 7.855 for the first, second and third mode of the natural frequency, respectively. Also, $\rho_{s}^*$, $A^*$, and $L^*$ are the structure density, cross-sectional area, and length of the plate, respectively. Using eqs.~\ref{eq:1} and \ref{eq:2}, the non-dimensional form of eq.~\ref{eq:4} is expressed as follows,
\begin{equation} \label{eq:5}
f_{ni}=\frac{k_{i}^2}{2\pi L }\sqrt{\frac{K_{b}}{M}}.
\end{equation} 
\rb{Kundu et al. \cite{kundu2017response} also studied the effect of Reynolds number ($Re$) and length of the splitter plate on its flapping frequency.} Shukla et al. \cite{shukla2013dynamics} experimentally showed that the amplitude of the oscillation of a splitter plate attached to a circular cylinder increases with $Re$ based on the cylinder diameter and they reported a plateau oscillation amplitude at $Re > 4000$. \rb{Very recently, Sahu et al. \cite{sahu_furquan_mittal_2019} numerically investigated of the dynamics of an elastic splitter plate mounted on an elastically-mounted cylinder.}

The above literature survey shows that the FIV of the plate exhibits complex and coupled physics, and most of the previous investigations \cite{connell2007flapping, liu2014stable, lee2014flapping, xu2016embedded, shoele2016flutter} neglected the thickness of the plate in the modeling and/or considered a membrane-like structure. \rb{For instance, the ratio of thickness to length, $K_b$ and $M$ were restricted to $O$(0.01), $O$($10^{-3}$) and $O$(1), respectively, in the previous reports. In the context of the development of FSI solver, while previous studies successfully demonstrated the advantage of Aitken's method including those in IB based solvers \cite{wall2008, BORAZJANI2, degroote2010performance, KIM2018296}, an FSI solver with a strongly coupled high-fidelity structural solver and that can handle large-scale FID of the structure has not been reported thus far, to the best of our knowledge.} 

Therefore, to address these computational challenges and capture the coupled physics during FIV, the objective of the present study is two-fold. The first is to develop and benchmark a high-fidelity FSI computational model, that can tackle the large-scale FID of a thin structure. In particular, the implicit \rb{(strong)} coupling between an in-house sharp-interface IB method based flow solver and an open-source, finite-element based structural dynamics solver is significantly accelerated using a dynamic under-relaxation method in the present work. \rb{To avoid the divergence of the coupling residual for challenging cases (e.g. low structure-fluid density ratio), additional sub-schemes have been implemented to bring robustness and numerical stability to the FSI coupling. For instance, we switch to constant under-relaxation if Aitken's method diverges.} Second, the present study also aims to generate new numerical data-sets while extending the FSI benchmark proposed by Turek and Hron \cite{turek2006proposal}, which could serve as additional benchmark data for future studies. These data-sets correspond to a wide range of mass ratio ($M$) and bending stiffness ($K_b$) of the plate. The second objective is to investigate the coupled dynamics of an elastic splitter plate attached to a cylinder, subjected to laminar flow. We consider a wide range of $M$ = [0.143, 20.029], $K_b$ = [0.0008, 0.0436] and $U_R$ = [2.562, 30.311], at $Re$ = 100, as compared to the previous reports. 

\section{Computational Model}\label{sec:com_model}

We employ an in-house FSI solver based on a sharp-interface IB method and was developed by Mittal and co-workers \cite{mittal2008versatile,seo2011sharp,mittal2011toward,bhardwaj2012benchmarking}. In the present work, the implicit coupling between the flow and structural dynamics solver has been implemented with dynamic under-relaxation \cite{Atul} and several code validations against FSI benchmarks are carried out \cite{Anup}. In the following subsections, different components of the solver are described briefly. The definitions of the major symbols used in the model are given in Table 1. 

\subsection{\textbf{Fluid dynamics}}\label{sec:code_validation}
The flow is governed by the \rb{two-dimensional}, unsteady, viscous, incompressible Navier-Stokes equations \rb{for a Newtonian fluid}, written in dimensionless form as follows,
\begin{equation} \label{eq:41}
\rb{\frac{\partial u_{i}}{\partial x_{i}}=0},
\end{equation}
\begin{equation} \label{eq:42}
\rb{\frac{\partial u_{i}}{\partial t}+u_{j}\frac{\partial u_{i}}{\partial x_{j}}=-\frac{\partial p}{\partial x_{i}}+\frac{1}{Re} \frac{\partial^2 u_{i} }{\partial x_{j} \partial x_{j} }},
\end{equation}
where\textit{ i}, \textit{j} = 1, 2, and $u_i$, $t$, $p$ and $Re$ are velocity components, time, pressure, and Reynolds number ($Re$) is based on mean flow velocity and cylinder diameter, respectively. The computational methodology to solve the above governing equations \rb{including discretization of the equations and algorithm of the flow solver} has been previously well-documented and details can be found in previous papers  \cite{mittal2008versatile,seo2011sharp,mittal2011toward}. 

To treat fluid-structure interface in the fluid domain, a sharp-interface IB method based on a multi-dimensional ghost-cell methodology developed by Mittal et al.\cite{mittal2008versatile} is employed. In this method, the governing equations for the fluid domain are solved on a non-uniform Cartesian grid in the Eulerian framework and the moving structure boundary is tracked within a Lagrangian framework. A schematic of the ghost-cell method is shown in Fig. 1(a). The cells whose centers are located inside the structure are identified as structure cells and the other cells outside the structure are identified as fluid cells. A structure cell which has at least one fluid cell as a neighbor is called a ghost-cell. A normal probe is extended from a ghost cell to intersect with the fluid-structure interface at a point, defined as body intercept point (Fig. 1(a)). The probe is extended into the fluid to the image point such that the body-intercept lies midway between the image and ghost points, as shown in Fig. 1(a). The kinematic boundary condition at the interface is prescribed by specifying an appropriate value at this ghost-cell. 

\rb{While tackling a moving fluid-structure interface, the sharp-interface IB methods are usually prone to spurious pressure oscillations due to the generation of ``fresh" and ``dead" cells \cite{mittal2008versatile}. The fresh (dead) cells are that fluid (solid) cells which were solid (fluid) cells in the previous time step. A cut-cell method proposed by Seo and Mittal \cite{seo2011sharp} is utilized to reduce the spurious pressure oscillations generated by fresh and dead cells.} 

\subsection{\textbf{ Structure dynamics}}\label{sec:SDS}
Here we briefly describe governing equation and constitutive model for the structure and more details are given in our previous paper \cite{bailoor2017fluid}. The Navier equations i.e., momentum balance equation in Lagrangian form are expressed as follows, 
\begin{equation} \label{(15)} 
\rho _{s} \frac{\partial ^{2} d_{i} }{\partial t^{2} } =\frac{\partial \sigma _{ij} }{\partial x_{j} } +\rho _{s} f_{i}\rb{,} 
\end{equation} 
where\textit{ i} and \textit{j} range from 1 to 3, \textit{$\rho$}${}_{s}$ is the structure density, \textit{d}${}_{i\ }$is the displacement component in the \textit{i} direction, \textit{t} is the time, \textit{$\sigma$}${}_{ij}$\textit{ }is the Cauchy stress tensor and \textit{f}${}_{i}$ is the body force component in the \textit{i} direction. The displacement vector \textbf{d}(\textbf{x}, \textit{t}) describes the motion of each point in the deformed structure as a function of space \textbf{x} and time \textit{t}. 

We employ Saint Venant-Kirchhoff material for the structure, that considers geometric non-linearity for a linear elastic material. The constitutive relation between the stress and the strain is based on Green-Lagrangian strain tensor $\mathbf{E}$ and second Piola-Kirchhoff stress tensor $\mathbf{S(E)}$ as a function of $\mathbf{E}$. The second Piola-Kirchhoff stress tensor can be expressed in terms of the Cauchy stress tensor $\mathbf{\sigma}$ as follows
\begin{equation}	\label{eq:IBM_SS5}
\mathbf{S} = J\mathbf{F}^{-1} \mathbf{\sigma} \mathbf{F}^{-T},
\end{equation}
where $J$ is the determinant of the deformation gradient tensor $\mathbf{F}$. The Green-Lagrangian strain tensor $\mathbf{E}$ is defined as follows,
\begin{equation}	\label{eq:IBM_SS6}
\mathbf{E} = \frac{1}{2}(\mathbf{F}^{T}\mathbf{F} -\rb{\mathbf{I}} ).
\end{equation}
The input parameters to the constitutive model are Young\textsc{\char13}s modulus ($E$) and Poisson ratio ($\nu$). The Navier equations are solved using Galerkin finite-element method, implemented in Tahoe, an open-source, Lagrangian, three-dimensional, finite-element solver \rb{(Tahoe was developed at Sandia National Labs, U.S.A.)}. The details of the numerical methodology has been documented in previous papers \cite{bhardwaj2014computational, bailoor2017fluid}. 

\subsection{ \textbf{Implicit coupling with dynamic under-relaxation}}
In order to couple the flow and structural solvers, the continuity of velocity on the fluid-structure interface \rb{i.e. no-slip condition} is applied for the fluid domain, \rb{expressed as follows,} 
\begin{equation} \label{(26)} 
u_{i,f} ={\mathop{d_{i,s} }\limits^{\bullet }}, 
\end{equation} 
\rb{where subscripts \textit{f} and \textit{s} denote the fluid and structure, respectively. The continuity of the traction is prescribed at the fluid-structure interface, given by,} 
\begin{equation} \label{(27)} 
\sigma _{ij, f} n_{j} =\sigma _{ij, s} n_{j}, 
\end{equation} 
\rb{where $n_j$ is local normal pointing outward on the fluid-structure interface in the fluid domain. The pressure on the interface is computed using interpolated pressure at the boundary intercept points via a trilinear interpolation (bilinear interpolation for 2D), as described by Mittal et al. \cite{mittal2008versatile}.} 

\rb{The flow and structural solvers are coupled using an implicit partitioned approach using a constant value of under-relaxation, as described by Bhardwaj and Mittal \cite{bhardwaj2012benchmarking}. In the present work, we implement dynamic under-relaxation factor ($\omega$), estimated using Aitken's method. In the implicit coupling, the flow solution is marched by one time step with the current deformed shape of the structure and the velocity of the fluid-structure interface act as the boundary condition in the flow solver, as shown in the flow chart in Fig. 1(b). The structural solver is marched by one time step with the updated fluid dynamic forces. The FSI convergence is declared if L${}_{2}$ norm of the displacement or velocity of the interface reduces below a preset value (Fig. 1(b)).} 

\rb{In Aitken's method, two previous FSI sub-iterations are used to predict a better value of $\omega$. The expression of $\omega$ at a given FSI sub-iteration $k$ is given by \cite{degroote2010performance},}
\begin{equation} \label{eq:IBM_24}
\rb{\omega_k = -\omega_{k-1}\frac{(\mathbf{R}_{k-1})^T(\mathbf{R}_k - \mathbf{R}_{k-1})}{||{{\mathbf{R}_k - \mathbf{R}_{k-1}}||^2}},}
\end{equation}
\rb{where $\mathbf{R}$ is the interface residual vector and is defined as follows,}
\begin{equation}\label{eq:IBM_24}
\rb{\mathbf{R} = {\mathbf{I}_{new}}-{\mathbf{I}_{old}},}
\end{equation}
\rb{where 
$\mathbf{I}$ is an interface variable namely, position ($\mathbf{X}$), velocity ($\mathbf{V}$) and acceleration ($\mathbf{A}$) of the interface. $\mathbf{R}$ is difference of new and old value of $\mathbf{I}$ in a sub-iteration. Each variable is composed of 2 components (3 in three-dimensional), thus, we could use 6 (9 in three-dimensional) different variables to compute $\omega$, using eq. 13. In the present work, we used velocity of the interface in $y$-direction to compute $\omega$.}

\begin{figure}
 \centering
 \includegraphics[width=13cm,keepaspectratio]{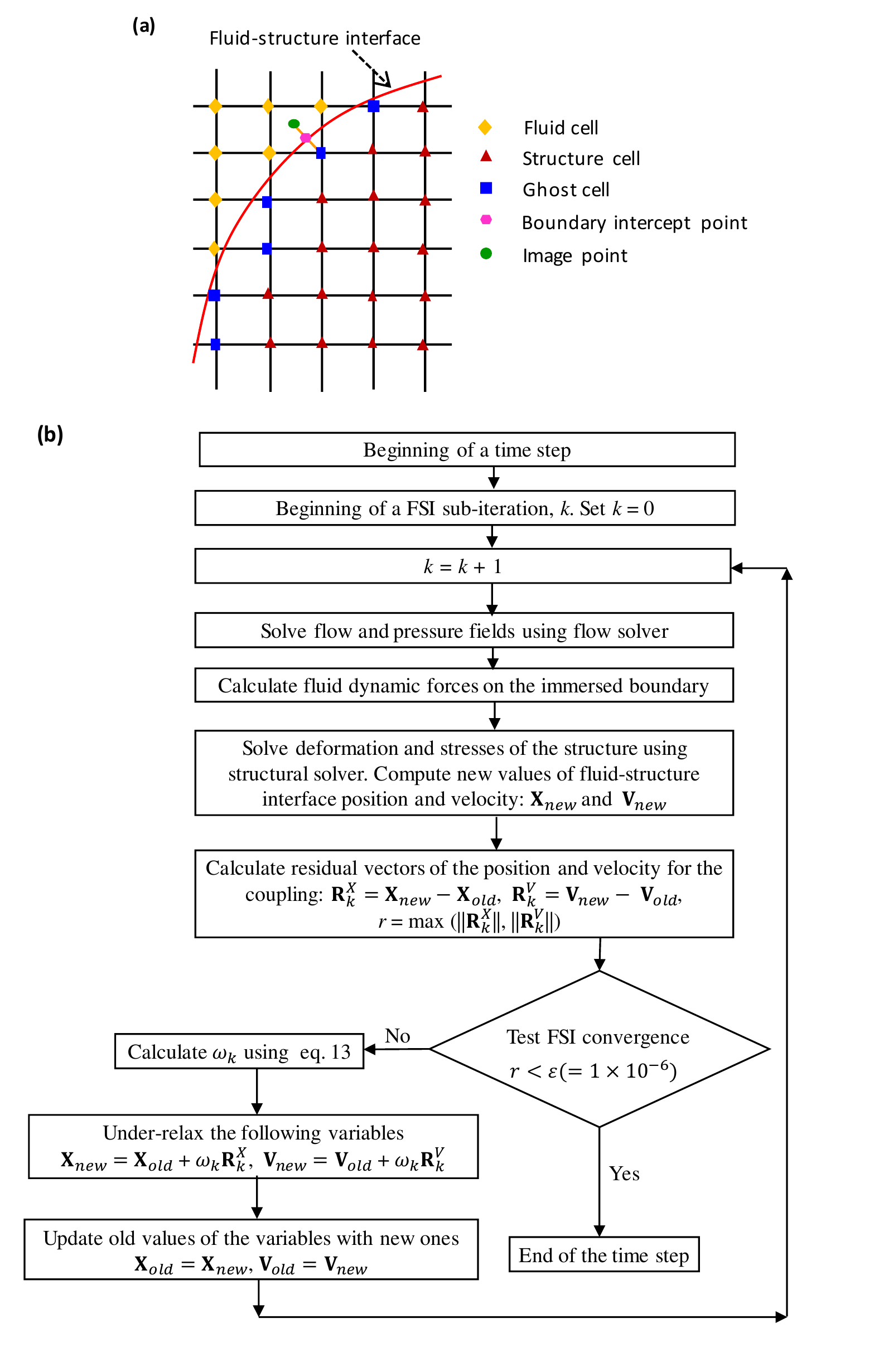}
 \caption{\rb{(a) Schematic of the IB method proposed by Mittal et al. \cite{mittal2008versatile} (b) Flow chart of the implicit coupling between flow and structural solver utilizing dynamic under-relaxation method for one time-step.}}
 \label{fig:flowchart}
\end{figure}

\rb{We start the FSI sub-iterations with a small initial guess of $\omega$ (say $\omega_0$). A better guess of $\omega_0$ is the value obtained using constant under-relaxation value, to achieve the FSI convergence.  As shown in the flow chart in Fig.~\ref{fig:flowchart}(b), we calculate FSI residual ($r$) based on $\mathbf{X}$ and $\mathbf{V}$ over successive sub-iterations and convergence is declared if the residual reduces below a defined threshold value. We compute $\omega$ at a given sub-iteration $k$ using eq. 13 and keep revising the position and velocity of the fluid-structure interface ($\mathbf{X}$ and $\mathbf{V}$) until the convergence is achieved in a given time-step.}

\rb{Aitken’s method is exact for linear systems, implying that 3 iterations (2 previous guesses plus 1 Aitken’s) are needed for the convergence of a linear system in a given time-step. However, due to the large non-linearity of the present FSI system, the coupling may diverge due to the large value of $\omega$, predicted by the Aitken’s method. To circumvent this problem and to ensure convergence, we implement the following two sub-schemes in the algorithm. First, the value of $\omega$ is restricted in a defined range, [$\omega_{min}$, $\omega_{max}$]. Thus, a large prediction of $\omega$ by the Aitken’s method is superseded by $\omega_{max}$. Second, in case of a divergence even with $\omega_{max}$, we switch to constant under-relaxation value for few sub-iterations, say $N_{safe}$ and use a smaller under-relaxation value, say $\omega_{safe}$. We switch back to the dynamic under-relaxation after $N_{safe}$ iterations. We have used $N_{safe}$ = 5 and $\omega_{safe} = \omega_0 = 0.1$ in the present work. The improvement in the FSI convergence as well as numerical stability for a test case is discussed in section 3.4.}  

\section{Benchmarking and testing of the FSI solver} \label{sec:results}

\subsection{\rb{Grid-size independence study for fluid domain}}\label{sec:griddomain} 

We examine grid-size convergence for the FSI benchmark problem, proposed by Turek and Hron \cite{turek2006proposal}, in an {\it{open domain}} instead of a channel. As shown schematically in Fig.~\ref{fig:validation}, we consider a thin, elastic splitter plate with dimensions 3.5$D$ $\times$ 0.2$D$, mounted at the lee side of a rigid cylinder of diameter, \rb{$D$}. The domain length and width are $(S_1 + S_2)$ and $S_3$, respectively, where $S_1$, $S_2$ and $S_3$ are taken as $5.5D$, $14D$ and $12D$, respectively. The center of the cylinder is ($S_1, S_3/2$). 

The boundary conditions for the present work are illustrated in Fig.~\ref{fig:validation}. \rb{Inflow and outflow velocity boundary conditions are prescribed at the left and right boundary, respectively. At left boundary, the following inlet velocity is prescribed, $u_i$ = (1,0). At the top and bottom boundaries of the domain, zero shear stress boundary condition is prescribed.} No-slip is applied on the fluid-structure interface. The following values are considered for the simulation setup, $D = 1$ and $Re = 100$, based on the cylinder diameter and uniform velocity at the inlet. The dimensionless Young\textsc{\char13}s modulus, structure to fluid density ratio and Poisson ratio are taken as 100, $1.4 \times 10^3$, 10 and 0.4, respectively. 

\rb{Five cases of non-uniform Cartesian grids in the fluid domain with the following points are considered for carrying out grid-size independence study: 193 $\times$65, 257 $\times$97, 321 $\times$161, 385 $\times$ 193 and 481 $\times$ 193. A high resolution of the grid is incorporated into the region where the plate movement is expected (Fig. 3(a)) and a non-uniform grid stretching is used from this region to the boundary. The minimum grid sizes in $x$ and $y$ directions are kept same, $\Delta x_{min} = \Delta y_{min}$, in each case and are listed in Table~\ref{table:nonlin}. The ratio of minimum grid size, $\zeta = \Delta x_{min}^k/\Delta x_{min}^{k+1}$, between successive meshes has also been listed in Table 2, where $\Delta x_{min}^k$ is the grid size of case $k$. $\zeta$ is around 1.3, expect for the coarsest mesh considered (for case 1 $\zeta$ = 1.67). The time step for these unsteady simulations is set to $\Delta t$ = 0.01, based on time-step independence study.}

\rb{Time-varying Y-displacements of the tip of the plate obtained for different grids are compared in Fig.~\ref{fig:grid_domain}(a). We note minor differences in the maximum tip displacement for different cases plotted in the inset. Computed amplitudes of the vibration of the plate ($A_{Ytip}$) for different grids are also listed in Table 2. To quantify the grid size convergence, L$_2$ norm of the errors in the amplitude of the displacement signal ($A_{Ytip}$) with respect to the finest grid are estimated and are tabulated in Table~\ref{table:nonlin}. The L$_2$ error norms are plotted against the grid size ($\Delta x_{min} = \Delta y_{min}$) in Fig.~\ref{fig:grid_domain}(b) on a log-log scale. The errors approximately reduce along a line of slope 2, implying a second-order accuracy of the coupled FSI solver. The error in the case of 385 $\times$ 193 grid is one order of magnitude lesser than the coarsest grid considered (Table 2). Therefore, 385 $\times$ 193 non-uniform Cartesian grid with $\Delta x_{min}= 0.02$ and $\Delta y_{min}= 0.02$ was selected for all the simulations presented in section 4. }

\subsection{\rb{Domain-size independence study}}\label{sec:griddomain2} 

\rb{In order to test domain size independence, we vary the size, $(S_1+S_2) \times S_3/2$ and consider four domains of sizes, 19.5$D$ $\times$ 12$D$, 30$D$ $\times$ 12$D$, 40$D$ $\times$ 12$D$ and 50$D$ $\times$ 20$D$. 
Other simulation parameters are kept the same as described in section 3.1. We kept $S_1 = 5.5D$ for all cases and varied $S_2$ and $S_3$ to study the domain size independence, as tabulated in Table 3. Simulated values of the amplitude of the Y-displacement ($A_{Ytip}$) obtained for all cases are listed in Table ~\ref{table:wes} and we note minor differences in the values of the $A_{Ytip}$. The percentage differences in $A_{Ytip}$ with respect to the biggest domain considered (50$D$ $\times$ 20$D$) for the rest of the domains are also listed in Table ~\ref{table:wes}. The magnitude of the percentage difference concerning the 19.5$D$ $\times$ 12$D$ domain is lesser than 1\%. Therefore, 19.5$D$ $\times$ 12$D$ domain is subsequently used for simulations presented in section 4.}

\subsection{\rb{Grid-size independence study for structure domain}}\label{sec:griddomain3} 

\rb{We perform structural grid-size convergence study in an open domain with a 385 $\times$193 grid in the fluid domain and consider four different grids composed of triangular finite elements in the plate (Fig.~\ref{fig:s_m} (b-e)). The numbers of finite triangular elements ($N$) in different structural grids considered are listed in Table~\ref{table:table43}. \rb{The boundary conditions, domain size, and simulation parameters are kept the same, as discussed in section 3.1.} The time-varying displacements ($Y_{tip}$) obtained for the different structural grids are compared in Fig.~\ref{fig:s_m_r}(a). The comparison shows minor differences in the maximum displacement, as shown in the inset of the figure. Computed values of $A_{Ytip}$ for all cases are also tabulated in Table 4. In order to quantify the grid convergence, L$_2$ norms of the error with respect to the finest grid ($N$ = 3082) are listed in  Table~\ref{table:table43} and are plotted against $N$ on a log-log scale in Fig.~\ref{fig:s_m_r}(b). The errors approximately decay along a line of slope 2, implying a second-order accuracy. Since the error for $N$ = 2182 is one order of magnitude smaller (Table~\ref{table:table43}) than the coarsest grid considered ($N$ = 736), we choose $N$ = 2182 grid for the simulations presented in section 4.}

\subsection{Testing of faster convergence by using dynamic under-relaxation}\label{sec:validation}

We test the improvement in the convergence of the implicit coupling by dynamic under-relaxation scheme, described in section 2.3. The test problem is chosen as FSI benchmark problem proposed by Turek and Hron \cite{turek2006proposal} in an {\it{channel}}, described in the previous section. \rb{We used velocity of the interface in $y$-direction to compute $\omega$ (eq. 13) for the dynamic under-relaxation and the tolerance for the convergence is set to $2 \times 10^{-4}$.} The value of $\omega$ at which the constant under-relaxation converges for this problem is around 0.1. \rb{In order to demonstrate the advantage offered by the dynamic under-relaxation, we compare the following three cases. (i) Constant under-relaxation, $\omega$ = 0.1; (ii) Dynamic under-relaxation, $\omega_{min}$ = 0.1 and $\omega_{max}$ = 0.4; and (iii) Dynamic under-relaxation with $\omega_{min}$ = 0.05 and $\omega_{max}$ = 0.8.} 

\rb{Computed $y$-displacement of the tip of the plate is shown in Fig.~\ref{fig:IBM_7}(a). The plate attains a self-sustained periodic oscillation with a constant amplitude and frequency after $t$ $\approx$ 60. We examine the FSI convergence history at four time-instances, $t_1$ to $t_4$ (shown by black dots in Fig.~\ref{fig:IBM_7} (a)), in a cycle of plate oscillation. The variation of the FSI residual with sub-iterations is compared for constant and dynamic under-relaxation schemes for the time-instances in Fig.~\ref{fig:IBM_7}(b-e). It is noted that residual decreases much faster with dynamic $\omega$ in cases (ii) and (iii) as compared to constant under-relaxation, case (i), for all time-instances considered. The better performances of the latter two cases are almost similar at all instances in a typical cycle of the oscillation. The computed average numbers of iterations for one cycle of plate oscillation in cases (i), (ii) and (iii) are around 76, 39 and 28, respectively. This shows a reduction of around 3 and 2 times in the number of sub-iterations in cases (ii) and (iii), respectively, as compared to the case (i). On comparing cases (ii) and (iii), we note that the range of $\omega$ considered also influences the number of sub-iterations in the dynamic under-relaxation method. A linear variation of the residual (say between 15 to 20 sub-iterations at $t_1$) shows that the method uses a constant value of the under-relaxation, implying that the coupling scheme avoids the possible divergence. Overall, we demonstrate a reduction in FSI sub-iterations by three times using the dynamic under-relaxation scheme. the revised method ensures better numerical stability of the FSI solver at low structure-fluid density ratio and accelerates the convergence of the implicit coupling.}

\subsection{Code validations}\label{sec:validation}

The present FSI solver has been extensively validated in previous studies. The flow solver was validated by Mittal et al. \cite{mittal2008versatile} for benchmark CFD problems such as the flow past a circular cylinder, sphere, airfoil, suddenly accelerated normal plate and suddenly accelerated circular cylinder. \rb{Further, Kundu et al. \cite{kundu2017response} validated the flow solver for pulsatile inflow past a cylinder in a channel.} Large-scale FID of a thin, elastic splitter plate was validated against the FSI benchmark problem proposed by Turek and Hron \cite{turek2006proposal} by Bhardwaj and Mittal \cite{bhardwaj2012benchmarking} and Kundu et al. \cite{kundu2017response}. Recently, the code was validated for vortex-induced vibration of a circular cylinder \cite{garg2018sharp} and the FID of a viscoelastic splitter plate \cite{MISHRA2019284}. In the following sub-sections, first, we present validations of the large-scale FID module of the in-house solver, presented in section 2, for the FSI benchmarks proposed by Turek and Hron \cite{turek2006proposal} and Wall and Ramm \cite{wall1998fluid}. We used grid-size established in sections 3.1 and 3.3 for the fluid and structure domain, respectively. 

\subsubsection{Splitter plate attached to the circular cylinder} \label{sec:TH}

\begin{figure}
 \centering
 \includegraphics[width=15cm,keepaspectratio]{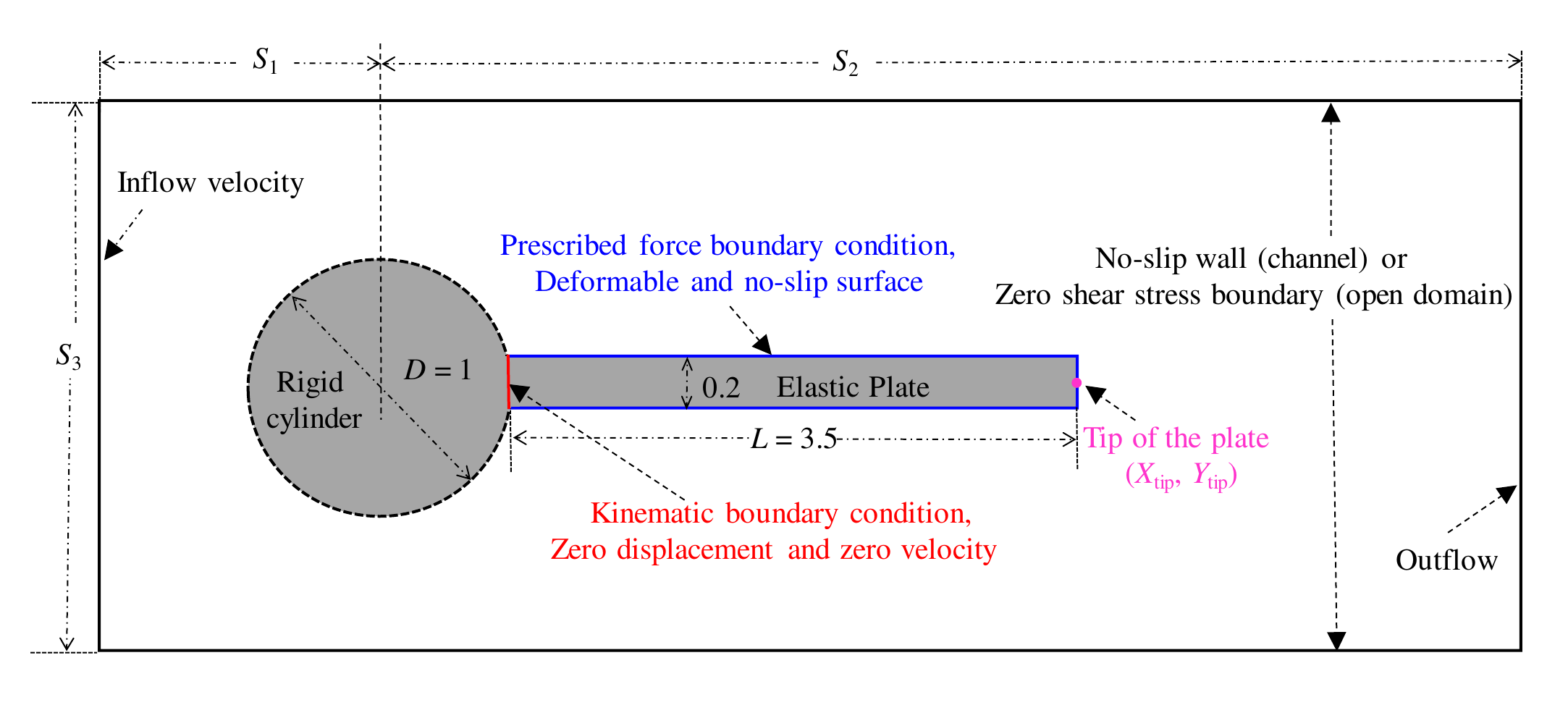}
 \caption{\rb{Schematic of computational domain with details of the boundary conditions considered for the validation with FSI benchmark proposed by Turek and Hron \cite{turek2006proposal} and for analyzing FIV of the plate in an open domain.}}
 \label{fig:validation}
\end{figure}

\begin{figure}
\centering
\includegraphics[width = 1.0\textwidth]{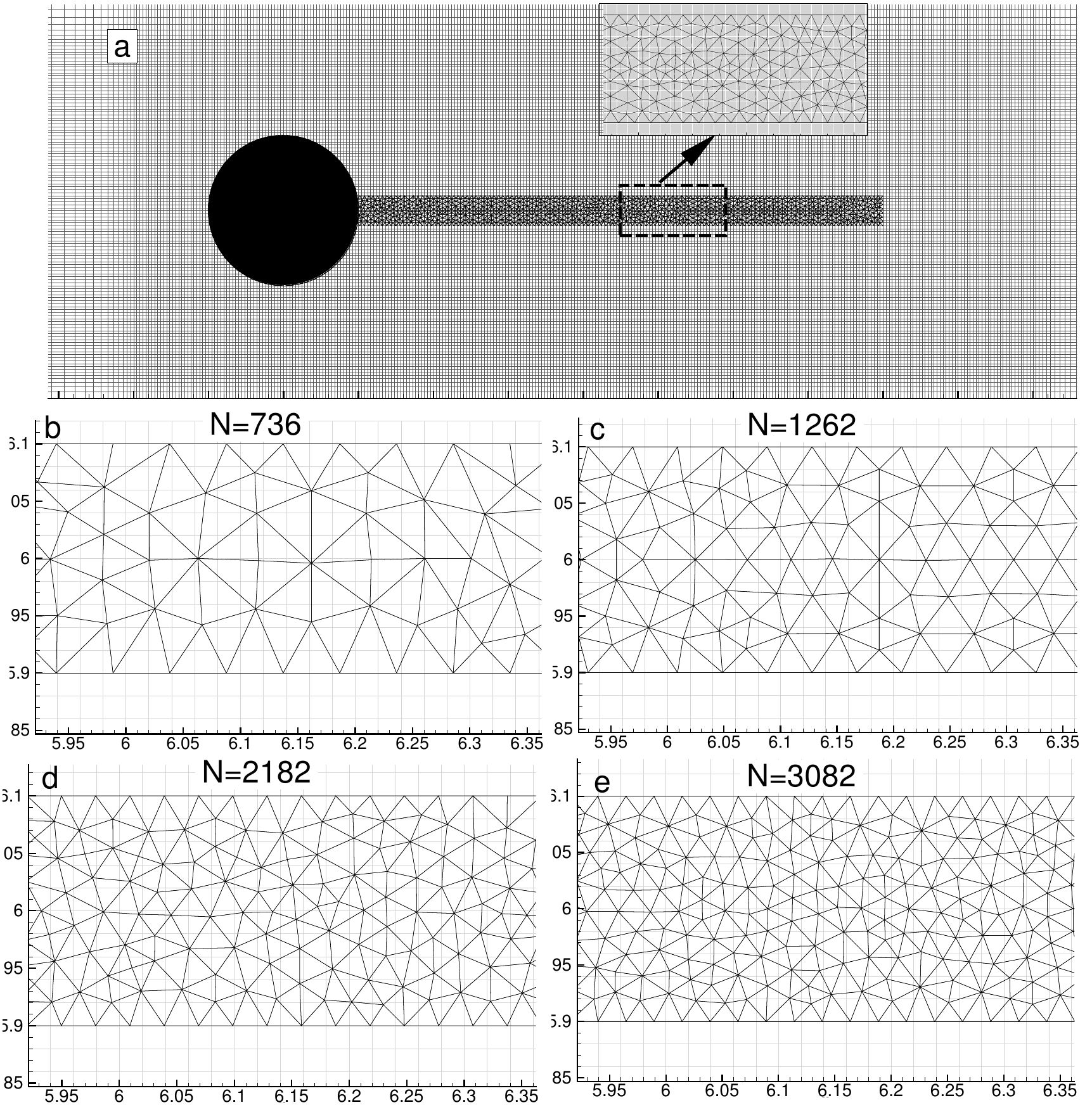}
\caption{\label{fig:s_m}\rb{(a) Finite-element grid of the plate immersed in a non-uniform Cartesian grid for the fluid domain. A uniform grid is used in the region in the fluid domain in which the plate is expected to move and non-uniform grid stretching is used from this region to the boundaries of the computational domain. Inset shows zoomed-in view of the finite-element grid in the plate. (b-e) Zoomed-in view of the finite-element grid for different cases of the grids considered in the plate, for carrying out grid-size resolution study. The number of finite-elements ($N$) is specified for each case.}}
\end{figure}

\begin{figure}
 \centering
 \includegraphics[width= 1.2 \textwidth]{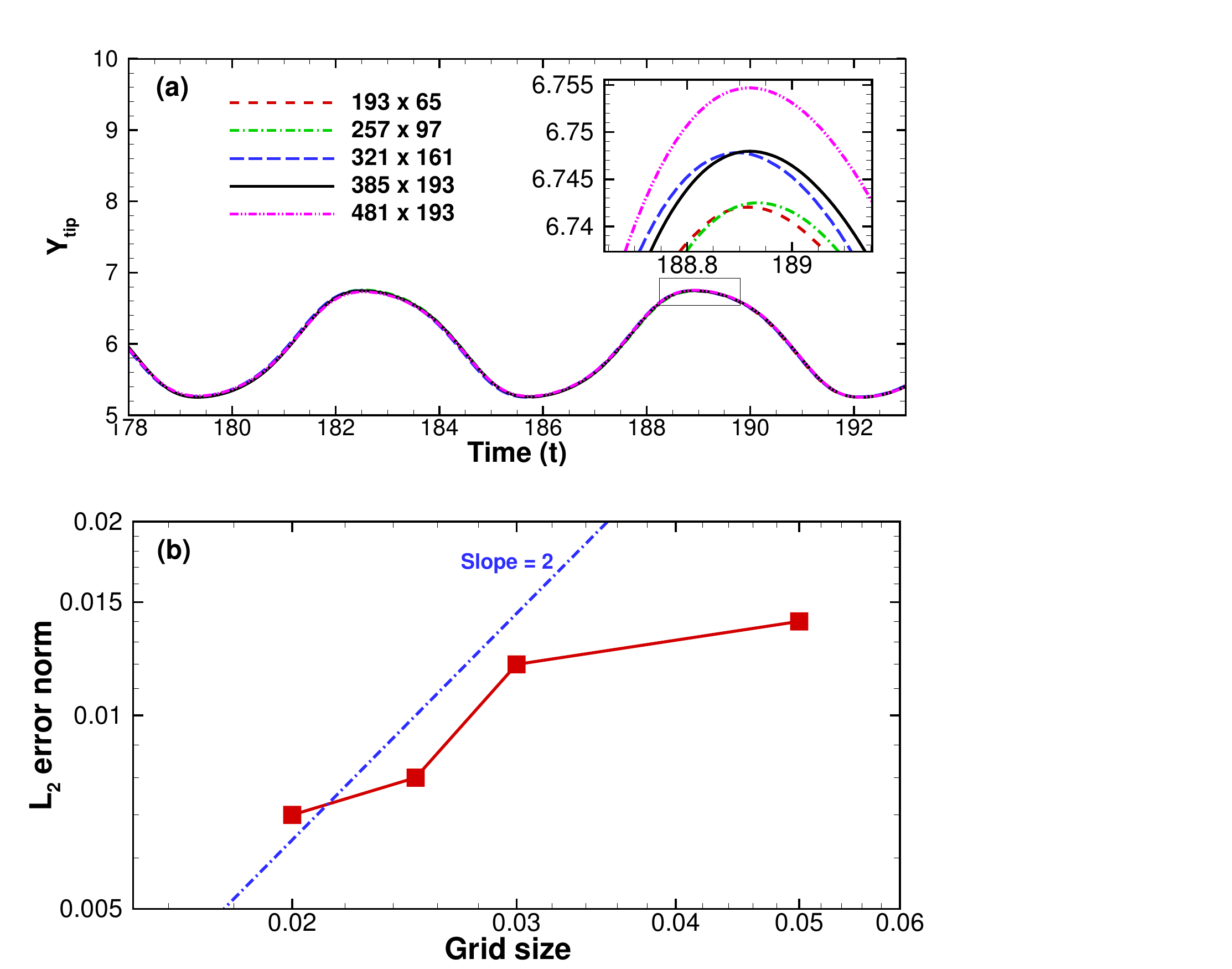}
 \caption{\rb{Grid size convergence study for the fluid domain (a) Comparison of the time-varying cross-stream displacement of the plate tip ( $Y_{tip}$) as a function of different grid resolution used in the fluid domain. (b) Variation of computed L$_2$ error norm of $A_{Ytip}$ of the grids considered against the grid size ($\Delta x_{min}= \Delta y_{min}$). The errors are computed with respect to the finest grid examined, 481 $\times$ 193 corresponding to $\Delta x_{min} = 0.015$.}}
 \label{fig:grid_domain}
 \end{figure}

\begin{figure}
\centering
\includegraphics[width = 1.2\textwidth]{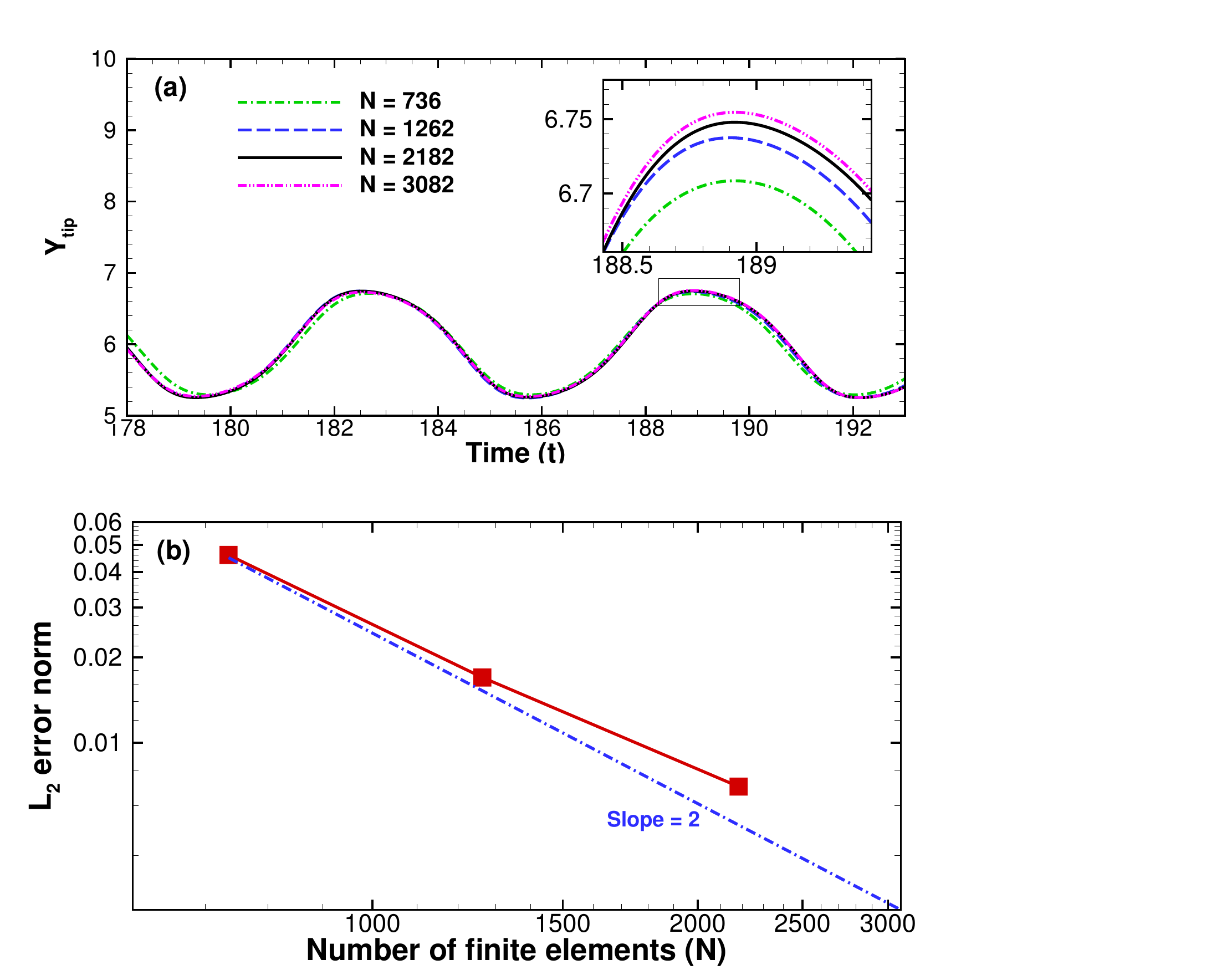}
\caption{\label{fig:s_m_r} \rb{Grid size convergence study for the structure domain. (a) Comparison of the time-varying cross-stream displacement of the plate tip ($Y_{tip}$) for different grids considered in the structure domain. The Cartesian grid for the fluid domain is kept same for each case. (b) Variation of computed L$_2$ error norm of $A_{Ytip}$ of the grids considered against the number of finite elements ($N$) for the different grids. The errors are computed with respect to the finest grid examined, $N$ = 3082.}}
\end{figure}

\begin{figure}
\centering
\includegraphics[width=13cm,keepaspectratio]{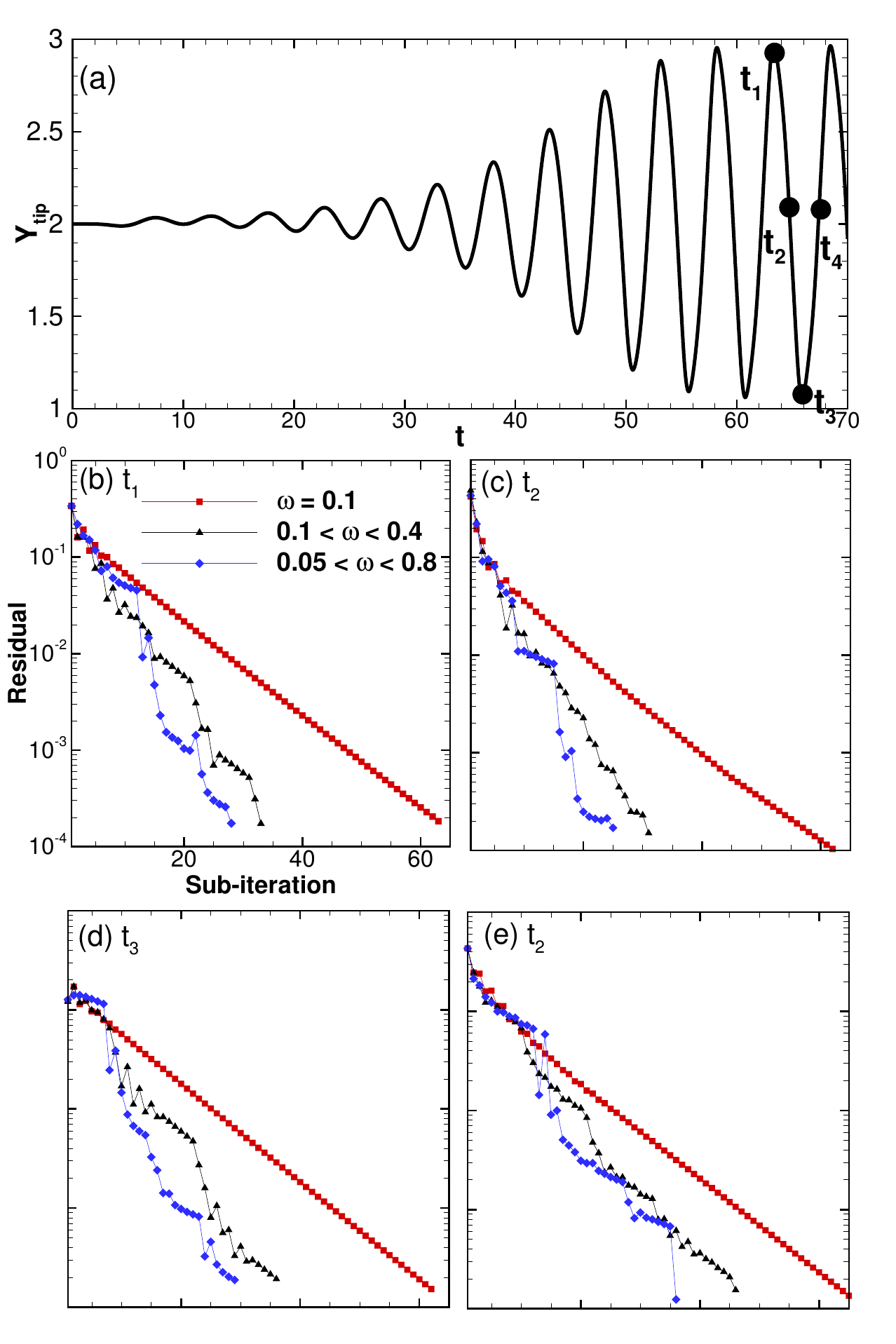}
\caption{\label{fig:IBM_7}\rb{Test of faster FSI convergence by dynamic under-relaxation. (a) Time-varying $Y$-displacement of tip of the plate for a typical oscillation cycle. (b-e) FSI residual variation with number of sub-iterations in the implicit coupling scheme at different time-instances for density ratio,  $\rho = {\rho_{s}}/{\rho_{f}}$ = 10. The time instances $t_1$, $t_2$, $t_3$ and $t_4$ are shown in (a). }}
\label{fig:fig6}
\end{figure}

We carried out the validation proposed by Turek and Hron \cite{turek2006proposal}, in which an elastic plate is mounted on the lee side of a rigid cylinder in a channel (Fig.~\ref{fig:validation}). The length and width of the channel are considered as $S_1 + S_2 = 20 D$ and $S_3 = 4.1 D$, respectively, shown in Fig.~\ref{fig:validation}. The center of the cylinder is (2$D$, 2$D$). The parabolic inflow boundary condition was imposed. The material parameters are taken as follows: Poisson ratio = 0.4, dimensionless Young\textsc{\char13}s modulus $(E)$ = 1400 and the structure to fluid density ratio of $\rho$ = 10. Based on the grid-size independence test presented in section 3.1, a non-uniform Cartesian mesh with 385 $\times$ 161 nodes was used for simulation with $ \Delta x_{min} $ = $ \Delta y_{min} $= 0.02 and non-dimensional time step of $ \Delta t $ = 0.01. 

\begin{figure}
 \centering
 \includegraphics[width=13cm,keepaspectratio]{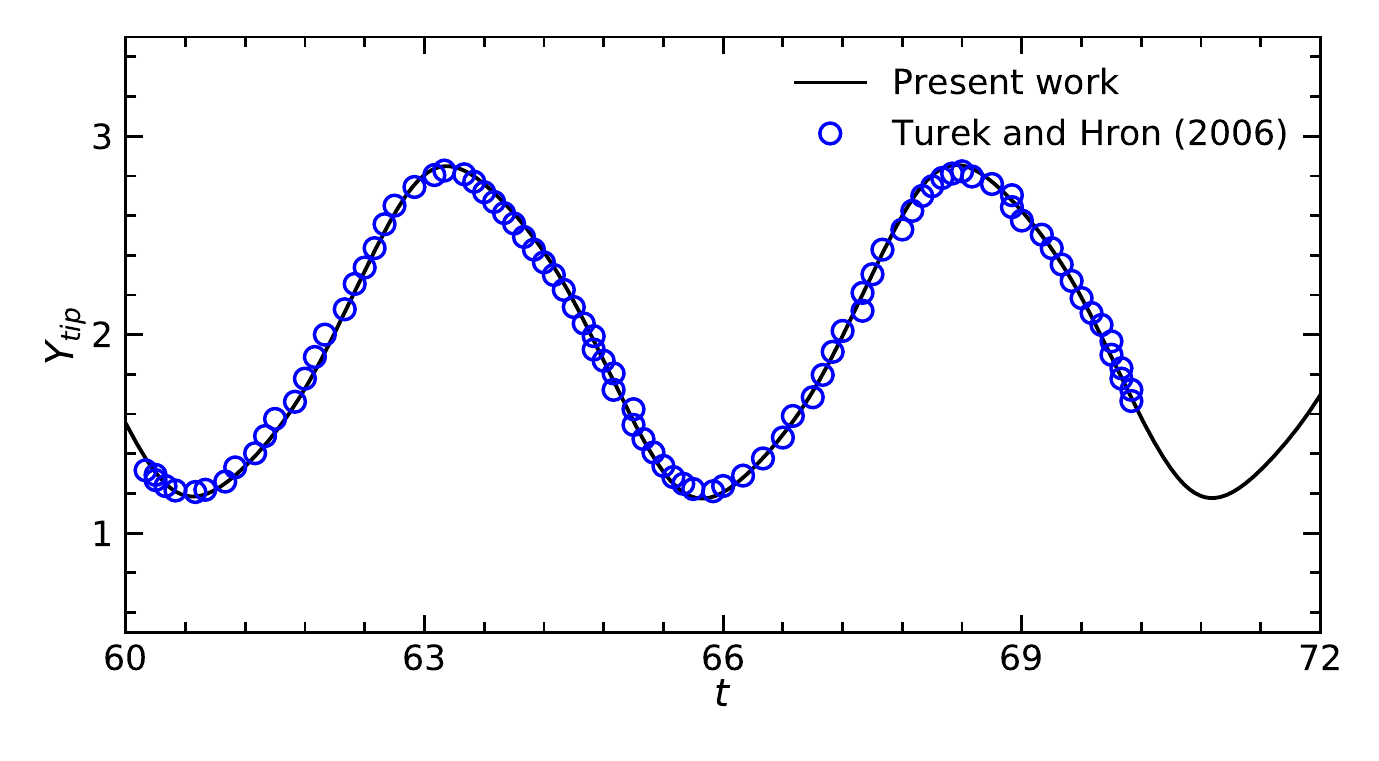}
\caption{Comparison of the computed tip displacement in $Y$-direction of the thin elastic splitter attached to the cylinder against the benchmark results of Turek and Hron \cite{turek2006proposal}.}
 \label{fig:validation1}
\end{figure}

The computed displacement of the tip of the plate is periodic and its amplitude reaches a plateau value \rb{at around $t$ $\approx$ 60 (Fig. 6(a))}. We compare the time-varying cross-stream position of the plate tip ($Y_{tip}$) with the benchmark data of Turek and Hron \cite{turek2006proposal} in Fig.~\ref{fig:validation1}. The amplitude ($A_{Ytip}$) and oscillation frequency ($f_p$) of the plate are in excellent agreement with the published data. We also extend the benchmark in an open domain, keeping all simulation parameters the same. \rb{Zero shear stress} boundary condition is applied at the top and bottom boundary (Fig.~\ref{fig:validation}) in this case. The plate displacement ( $Y_{tip}$) and its oscillation frequency ($f_p$) for the open domain are $0.75D$ and $0.155D/U$, which are slightly lower than the corresponding benchmark values for the channel. 

\subsubsection{Splitter plate attached to the square cylinder} \label{sec:problem_definition}

\begin{figure}
 \centering
 \includegraphics[width=\textwidth,keepaspectratio]{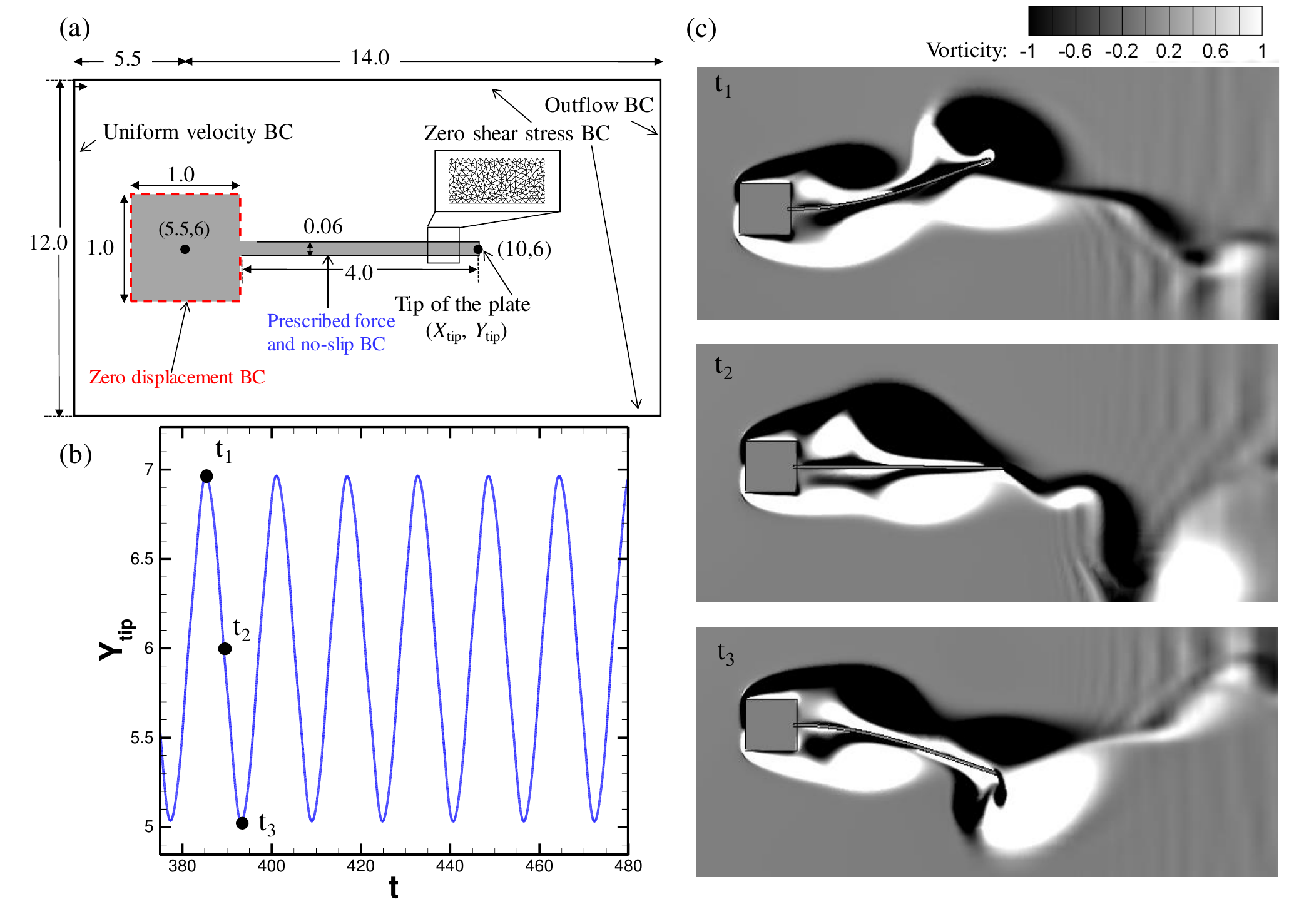}
 \caption{\rb{(a) Schematic of FSI benchmark problem proposed by Wall and Ramm \cite{wall1998fluid}. A thin elastic splitter plate is attached on the lee side of a square cylinder, subjected to laminar flow (b) Time-varying tip displacement in $Y$-direction after the plate reaches self-sustained oscillation. (c) Vorticity field at three different time-instances $t_1$, $t_2$ and $t_3$, as shown by black dots in (b).}}
 \label{fig:validation2}
\end{figure}

We further validate the large-scale FID module against the benchmark problem proposed by Wall and Ramm \cite{wall1998fluid}. In this problem, a thin elastic splitter plate is attached to a rigid square cylinder, as shown in Fig.~\ref{fig:validation2}(a). The reference length is taken as the side length of the square cylinder, and the reference velocity is taken as the inlet velocity. $Re$ based on these reference values is 333. The material parameters are taken as follows: Poisson ratio = 0.35, dimensionless Young\textsc{\char13}s modulus $E$ = 8.1 $\times$ $10^{5}$, and the structure to fluid density ratio $\rho$ = 84.7. The inlet flow conditions and boundary conditions are illustrated in Fig.~\ref{fig:validation2}(a). A non-uniform Cartesian mesh specified in the validation study has been used for this simulation, and the non-dimensional time step was set to $\Delta t $ = 7.5 $\times$ $10^{-3}$. The plate reaches a self-sustained periodic state, similar to the case of an elastic plate attached to a circular cylinder. The time history of the tip displacement ($Y_{tip}$) is plotted in Fig.~\ref{fig:validation2}(b). The computed plate vibration frequency, as well as the tip displacement along with published results, are listed in Table~\ref{table:square}. We found excellent agreement between the present and published results \cite{wall1998fluid, olivier2009fluid, habchi2013partitioned}, which further validates the present FSI solver. At the maximum tip displacement, contours of vorticity are shown at different time instances in Fig.~\ref{fig:validation2} (c). These time-instances are shown by dots in Fig.~\ref{fig:validation2}(b). As noted from the vorticity field, vortices shed alternatively at the top and bottom of the deforming plate.
 
\section{Analysis of FIV of an elastic splitter plate}

\rb{Numerical simulations were performed for the elastic splitter plate in an open domain using the same parameters, domain size and boundary conditions, as discussed in section 3.1.} The effects of mass ratio ($M$) and bending stiffness ($K_b$) on the dynamics of the elastic plate as a function of reduced velocity ($U_R$) are discussed in the following subsections. The definitions of these dimensionless variables are given in Table 1. 

\subsection{Effect of mass ratio ($M$)} \label{sec:M}

\begin{figure}
 \centering
 \includegraphics[width = 1.0 \textwidth]{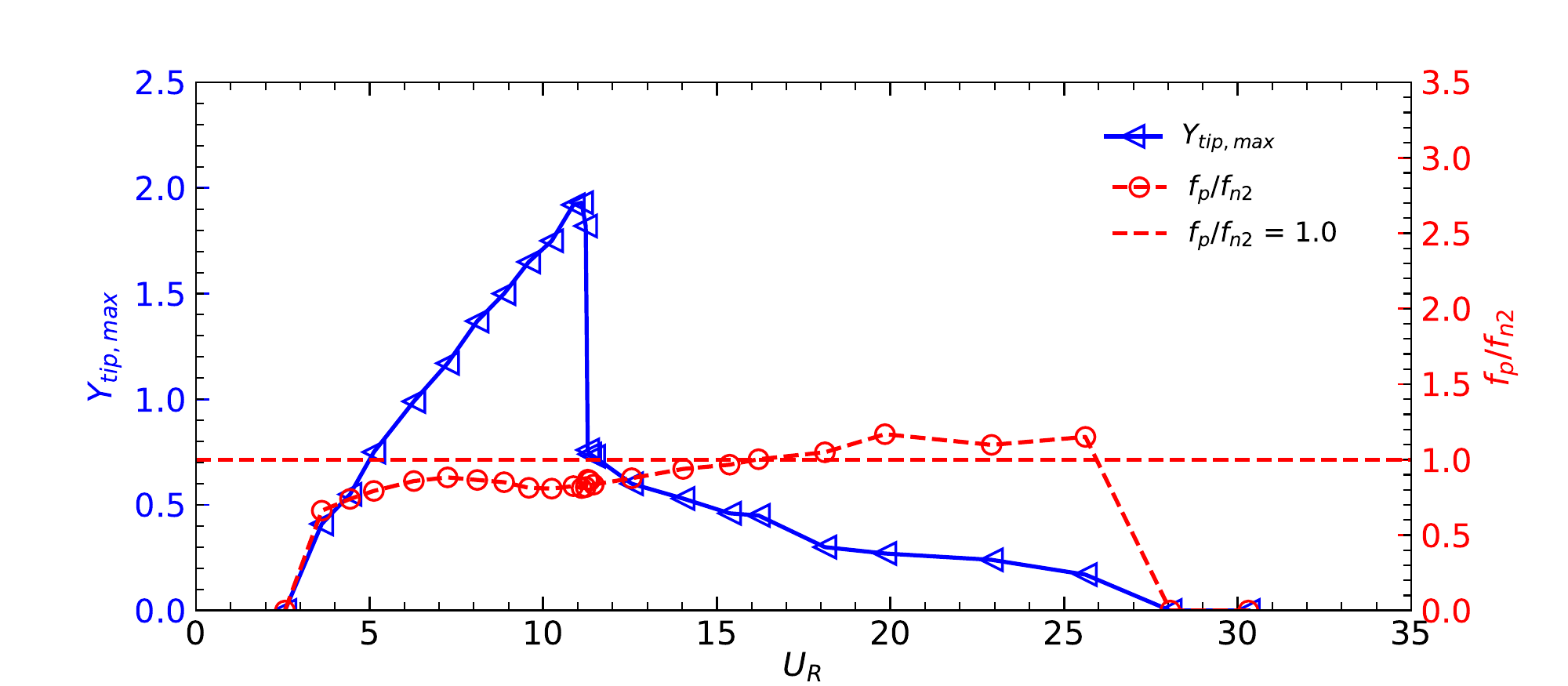}
 \caption{Computed values of the plate amplitude ( $Y_{tip,max}$) as a function of reduced velocity ($U_R$). The mass ratio was varied and bending stiffness is kept fixed at $K_{b} = 0.022$ for the cases plotted here. Ratio of plate oscillation frequency and second mode natural frequency of the plate in vacuum ($f_{p}/f_{n2}$) is plotted for all cases. A dotted line at $f_{p}/f_{n2}$ = 1 is also shown.}
 \label{fig:varym1}
\end{figure}

We discuss the effect of $M$ on the elastic splitter plate displacement ( $A_{Ytip}$), oscillation frequency ($f_p$) and wake structures, keeping $K_{b}$ constant. Simulations are presented for $M$ = [0.143, 20.029] and $K_{b} = 0.0218$, with simulation parameters given in Table~\ref{table:Mass}. Fig.~\ref{fig:varym1} plots amplitude of the tip of the plate, $A_{Ytip}$, as a function of $U_R$ for different cases of $M$. The range of $U_{R}$ considered is [2.562, 30.311] and $U_R$ increases with an increase in $M$, as noted in Table~\ref{table:Mass}. We also plot frequency ratio, $f_{p}/f_{ni}$ as a function of $U_R$, where $f_{p}$ and $f_{n2}$ are the frequencies of the plate and its natural frequency of oscillation in second mode, assuming it as a Euler-Bernoulli beam. These frequencies are obtained by the numerical simulation and eq. 5, respectively. 

Fig.~\ref{fig:varym1} shows that $A_{Ytip}$ increases with $U_R$ for $0.286 \leq M \leq 2.747$, reaching a maximum value of 1.93 at $M$ = 2.747 and then decreases with $U_R$ in the range $2.747 \leq M \leq 14.306$. The frequency ratio increases with $M$ (or $U_R$) for $0.286 \leq M \leq 1.717$ and is around 0.8-0.9 i.e., closer to unity for $1.717 < M < 3.434$. Fig.~\ref{fig:varym1} shows that a larger amplitude oscillation occurs for the cases where $f_{p}$/$f_{n2}$ $\approx$ 0.8-0.9. This is called as lock-in condition \rb{(discussed in detail in the next section)} and we plot a dotted line to denote the lock-in condition where $f_{p}$/$f_{n2}$ = 1. The curve starts to deviate from the dotted line over the mass ratio range $3.434 \leq M \leq 5.723$. \rb{At $M \geq 17.167$, $A_{Ytip}$ reduces to zero.} 

\subsubsection{Lock-in condition}

In the case of vortex-induced vibration (VIV) of an elastically mounted cylinder, the lock-in occurs if the vortex shedding frequency changes (and locks) to match the natural frequency of spring in vacuum \cite{Williamson2004} and the cylinder oscillates with a larger amplitude. In the present case, the simulated flow over a {\it{rigid}} splitter plate of length $L/D$ = 3.5 does not show any vortex shedding at $Re$ = 100. In the case of a rigid and sufficiently long splitter plate mounted on the lee side of the cylinder, the plate inhibits flow instabilities in the wake, and vortex shedding is suppressed. However, in the case of a shorter rigid plate, the shedding may occur. Therefore, we extend the definition of the lock-in in a classical VIV of a cylinder to the present FSI system, i.e. a rigid cylinder with an elastic splitter plate. In the latter, vortex shedding frequency of the FSI system and plate oscillation frequency ($f_{p}$) are identical and lock-in is defined if the vortex shedding or plate oscillation frequency ($f_{p}$) is close to the natural frequency of the plate in vacuum for any mode ($f_{ni}$), where subscript $i$ denotes the $i^{th}$ natural mode of the oscillation. Consequently, the plate oscillates with a larger amplitude in this condition.

We note similar characteristics for the FIV of the plate as described in the literature for the VIV of the cylinder. As discussed earlier, we note that large-amplitude oscillations for $1.71 \leq M \leq 3.43$ if $f_{p}$/$f_{n2}$ $\approx$ 0.8-0.9, indicating a lock-in region. The deviation of the frequency ratio from unity is attributed to added mass effect at low mass ratios \cite{khalak1997fluid, khalak1999motions}. In addition, a large reduction in $A_{Ytip}$ from $M$ = 2.74 to $M$ = 2.857 in Fig.~\ref{fig:varym1} is a typical transition from "upper branch" to "lower branch", as reported for the \rb{VIV of an elastically-mounted cylinder, constrained to move transverse to the flow \cite{Williamson2004}}. 

\subsubsection{Comparison of vibration characteristics of the plate and vorticity field}

\begin{figure}
 \centering
 \includegraphics[width=1.0\textwidth]{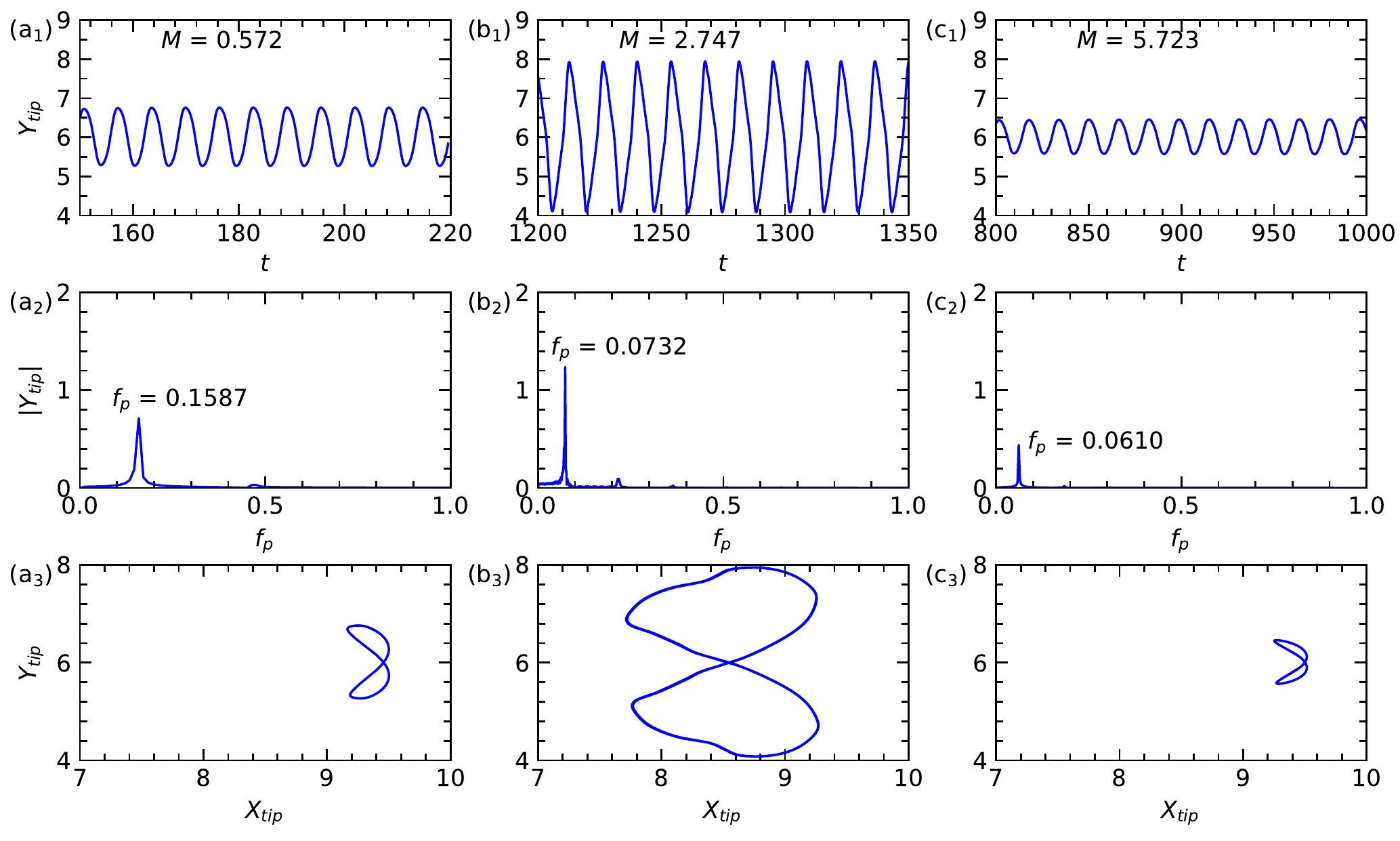}
\caption{Comparison of tip displacement $Y_{tip}$ (first row), power spectra (second row), and phase-plane plots (third row) for three cases of mass ratios, $M$ = 0.572, 2.747 and 5.723. The bending stiffness is kept same, $K_{b}= 0.0218$. Phase-plane plots are shown after the plate amplitude reaches a plateau value.}
\label{fig:ytipfft}
\end{figure}

Fig.~\ref{fig:ytipfft} (first row) compares the time-varying $Y_{tip}$ for mass ratios 0.572, 2.747 and 5.723. These cases are case 4, case 13 and case 20 of Table~\ref{table:Mass}. These signals are plotted after the plate reaches a self-sustained oscillation state. The computed values of the maximum plate amplitude values are 0.75, 1.93 and 0.48, respectively. Fig.~\ref{fig:ytipfft} (second row) shows the FFT of $Y_{tip}$ for the three cases. The dominant $f_{p}$ for mass ratios 0.572, 2.747 and 5.723 are 0.159, 0.073 and 0.061, respectively. Fig.~\ref{fig:ytipfft} (second row) shows that the plate with largest deformation for $M$ = 2.747 vibrates with more than one frequency. One frequency component is closer to a dominant second mode, and the other is a third harmonic of the dominant frequency. The phase plots in the third row of Fig.~\ref{fig:ytipfft} show the axial and lateral movement of the plate tip, which are larger for $M$ = 2.747 than other mass ratios because of the lock-in. As a result, the phase plots are wider at $M$ = 2.747 and each case exhibits a limit-cycle flapping \cite{connell2007flapping}. 

\begin{figure}
 \centering
 \includegraphics[width=1.0\textwidth]{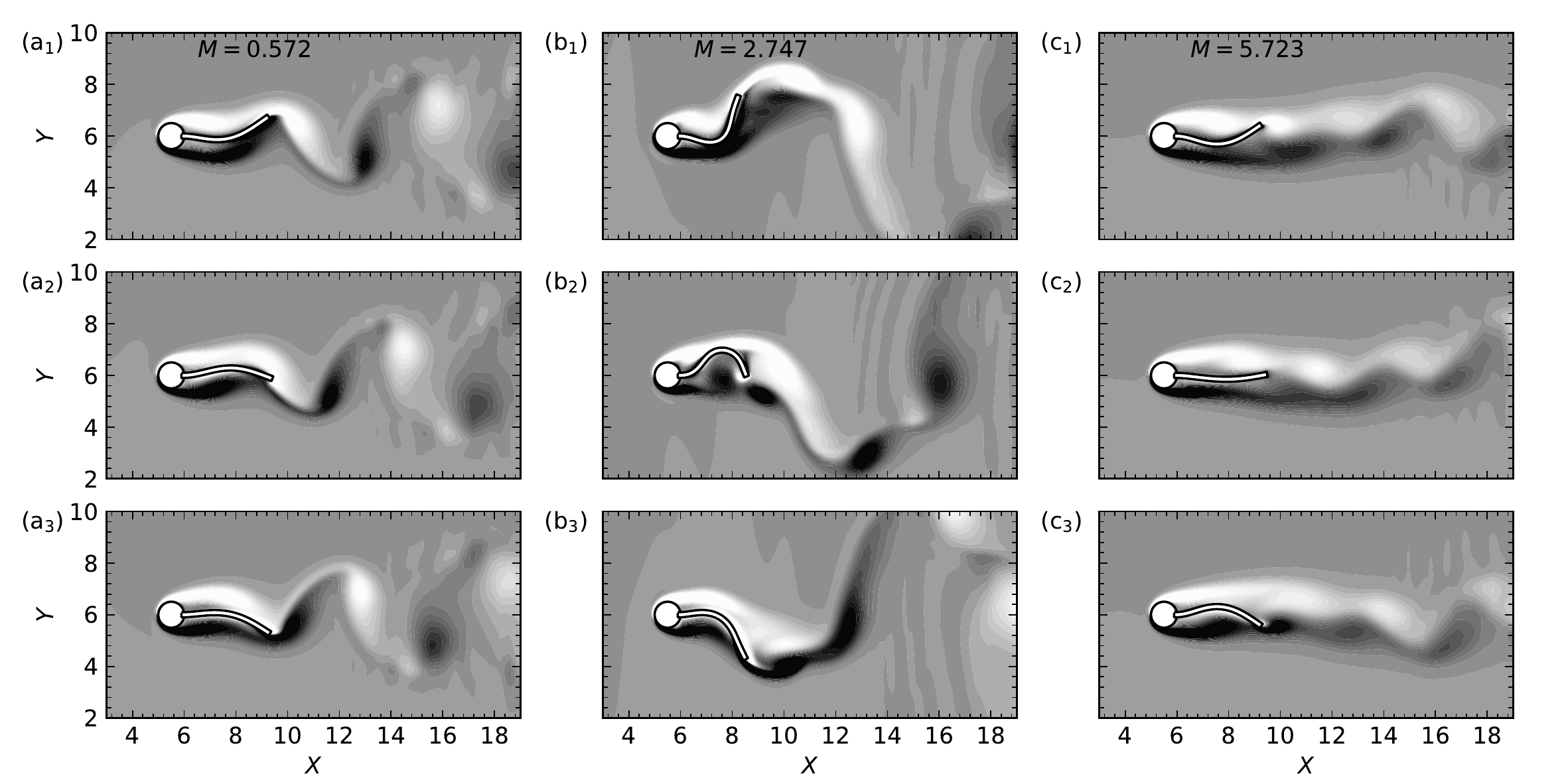}
 \centering
 \caption{\rb{Comparison of vorticity field at different time instances for three different mass ratios $M$ = 0.572, 2.747 and 5.723. The bending stiffness is kept same, $K_{b}= 0.0218$. The color map range is [-2, 2].}}
 \label{fig:vortexmass1}
\end{figure}

Fig.~\ref{fig:vortexmass1} depicts the instantaneous vorticity field around the plate. The vorticity is plotted in Fig.~\ref{fig:vortexmass1} at three different instances, corresponding to maximum, minimum and central positions of the plate. For $M$ = 0.572, the elastic splitter plate is in its self-sustained oscillation state and vortices are shedded alternately (see the first column of Fig.~\ref{fig:vortexmass1}). The vortex shedding shows "2S" vortex pattern, i.e., one vortex sheds from each side of the plate in a cycle \cite{govardhan2000modes} in the downstream. For $M = 2.747$ , the elastic plate bends to a greater degree compared to the $M$ = 0.572. As a result, longer vortex shedding from the elastic plate is observed, and it tends to split into two small vortices, which are not completely separated immediately. This vortex shedding flow pattern is typically similar to the 2P mode, i.e. two vortices shed from each side of the plate each time \cite{govardhan2000modes}. For $M = 5.723$, two positive and two negative vortices are shedding alternatively, as seen in the third column of Fig.~\ref{fig:vortexmass1}. Therefore, the vortex-shedding pattern depends on the FIV of the plate. 
 
\subsection{Effect of bending stiffness ($K_{b}$)}\label{sec:young} 

\begin{figure}
 \centering
 \includegraphics[width=1 \textwidth]{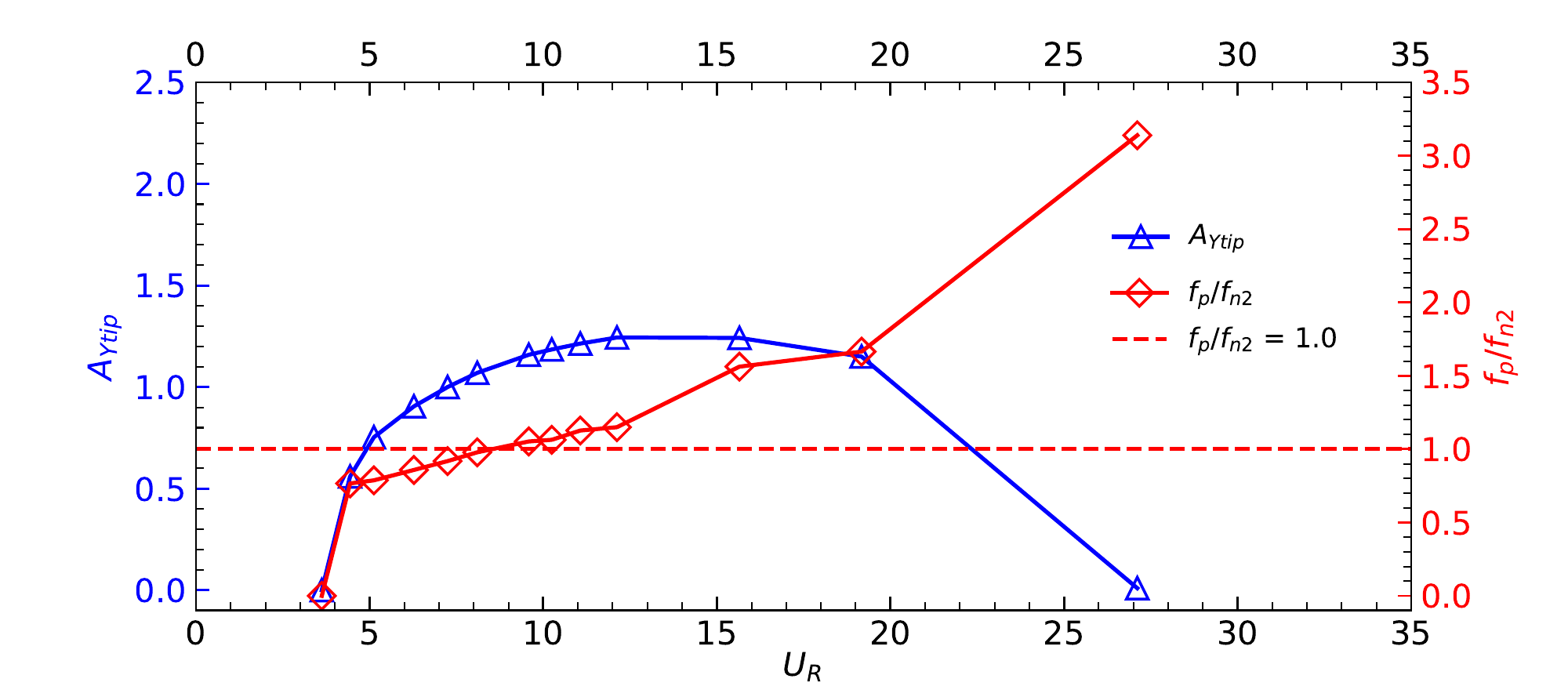}
 \caption{Computed values of the plate amplitude ( $A_{Ytip}$) as a function of reduced velocity ($U_R$). The bending stiffness ($K_b$) was varied and mass ratio is kept fixed at $M = 0.572$ for the cases plotted here. Ratio of plate oscillation frequency and second mode natural frequency of the plate in vacuum ($f_{p}/f_{n2}$) is plotted for all cases. A dotted line at $f_{p}/f_{n2}$ = 1.0 is also shown.}
 \label{fig:vary_Kb}
\end{figure}

The effect of $K_{b}$ on the FIV of the elastic splitter plate is examined in this section. We consider $K_{b} = [0.0008, 0.0436]$ keeping the mass ratio $M$ = 0.572 constant. The parameters of these simulation cases are given in Table~\ref{table:kb}. As done previously in section 4.1, we plot $A_{Ytip}$ and $f_{p}$/$f_{n2}$ as function of $U_R$ for different cases of $K_b$ in Fig.~\ref{fig:vary_Kb}. Note that $U_R$ decreases with an increase in $K_{b}$, as noted in Table~\ref{table:kb}. Fig.~\ref{fig:vary_Kb} shows that $A_{Ytip}$ is negligible at large $K_{b}$ (small $U_R$) and it increases with $U_R$ or decreases with $K_{b}$. The frequency ratio, $f_{p}$/$f_{n2}$, is plotted as a function of $U_R$ (or $K_{b}$) and a red dotted line corresponds to ratio 1 i.e., lock-in condition. A larger amplitude occurs if $f_{p}$/$f_{n2}$ is in range of [0.9, 1.6]. \rb{The deviation of the frequency ratio from unity ($f_{p}$/$f_{n2}$ ${\approx}$ 1.6) is attributed to added mass effect at low mass ratios \cite{khalak1997fluid, khalak1999motions, Williamson2004}.} The plate amplitude is almost negligible for very large or very small $K_{b}$. \rb{The plate exhibits higher modes of the vibration along with the second natural mode at larger $U_R$ (at $U_R$ = 15.652 and 19.170, Fig.~\ref{fig:vary_Kb}(b$_2$)) and desynchronizes with the wake at $U_R$ = 27.111.} In general, the plate exhibits similar characteristics, as seen for \rb{VIV of an elastically-mounted cylinder, constrained to move transverse to the flow \cite{Williamson2004}.}

\begin{figure}
 \centering
 \includegraphics[width=1 \textwidth]{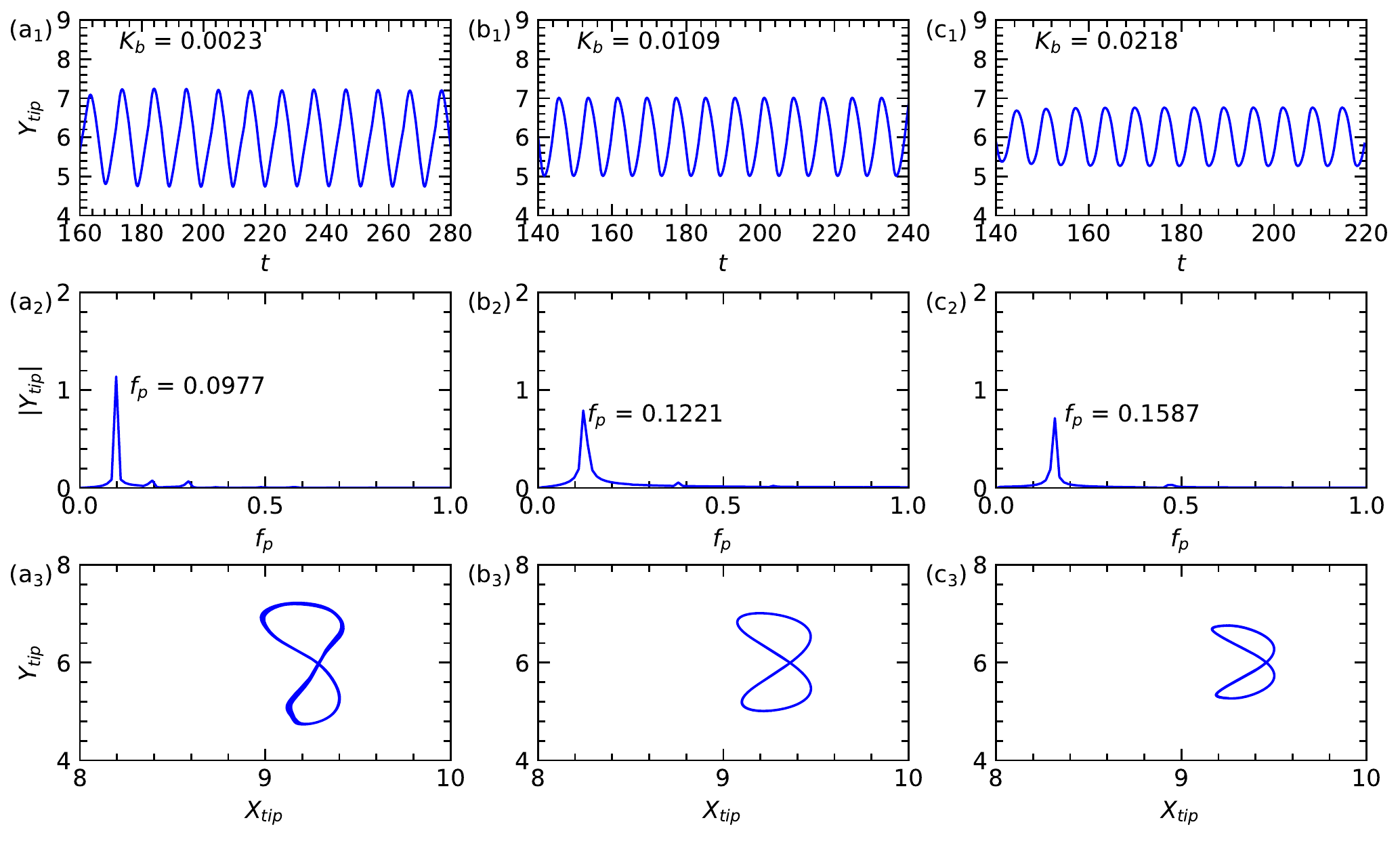}
 \caption{Comparison of tip displacement $Y_{tip}$ (first row), power spectra (second row), and phase-plane plots (third row) for three cases of bending stiffness, $K_b$ = 0.0023, 0.0109 and 0.0218, with same mass ratio, $M= 0.572$. Phase-plane plots are shown after the plate amplitude reaches a plateau value.}
 \label{fig:fult_Kbvary}
\end{figure}

\begin{figure}
 \centering
 \includegraphics[width=1.0\textwidth]{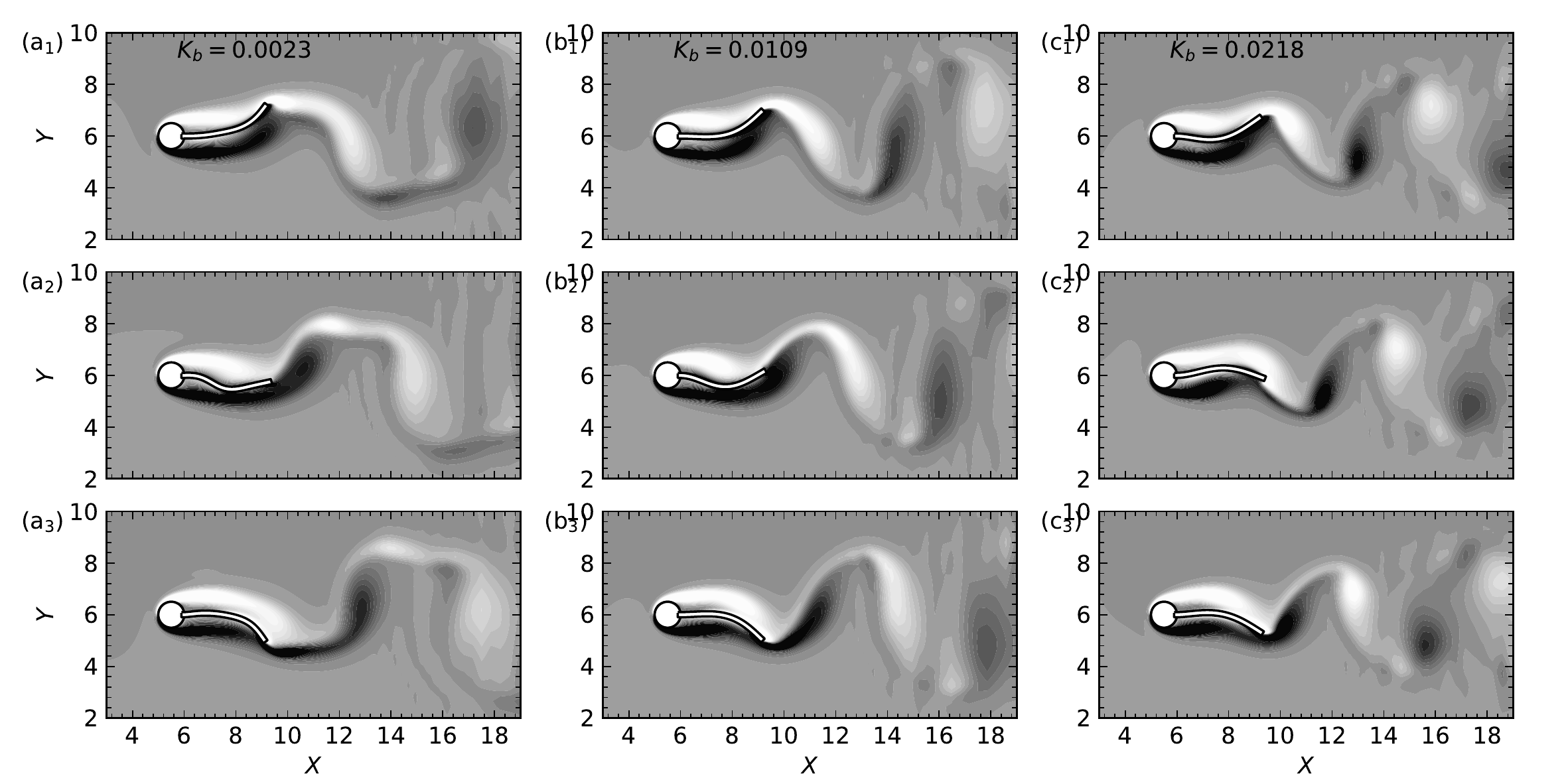}
 \caption{\rb{Comparison of vorticity field at different time instances for three different values of bending stiffness, $K_b$ = 0.0023, 0.0109 and 0.0218, with the same mass ratio, $M= 0.572$. The color map range is [-2, 2].}}
 \label{fig:Evari}
\end{figure}

Fig.~\ref{fig:fult_Kbvary} shows $Y_{tip}$, power spectra, phase plots of the plate in the first, second and third row, respectively. These three rows correspond to $K_b$ = 0.0023 (case 29 in Table~\ref{table:kb}), 0.0109 (case 35) and 0.0218 (case 4), respectively. The dominant frequencies plotted in Fig.~\ref{fig:fult_Kbvary} (second row) are closer to second mode natural frequency in the respective cases (e.g. $f_{n2}$ = 0.0530 $K_b$ = 0.0016, eq. 5). Fig.~\ref{fig:fult_Kbvary} (third row) shows that the phase plot is wider at $K_{b} =0.0023$ i.e., the plate shows larger bending as compared to other two cases. This is attributed to the lock-in condition for this case. Fig.~\ref{fig:Evari} shows the vorticity field for three cases of bending stiffness. Alternate vortices shed in the downstream in each case in 2S mode for all the cases \cite{govardhan2000modes}. The vortex structures become elongated at lower bending stiffness as compared to high bending stiffness due to the larger plate displacement.

\section{\rb{Applications of present study in energy-harvesting}}
\begin{figure}
 \centering
 \includegraphics[width=0.8\textwidth]{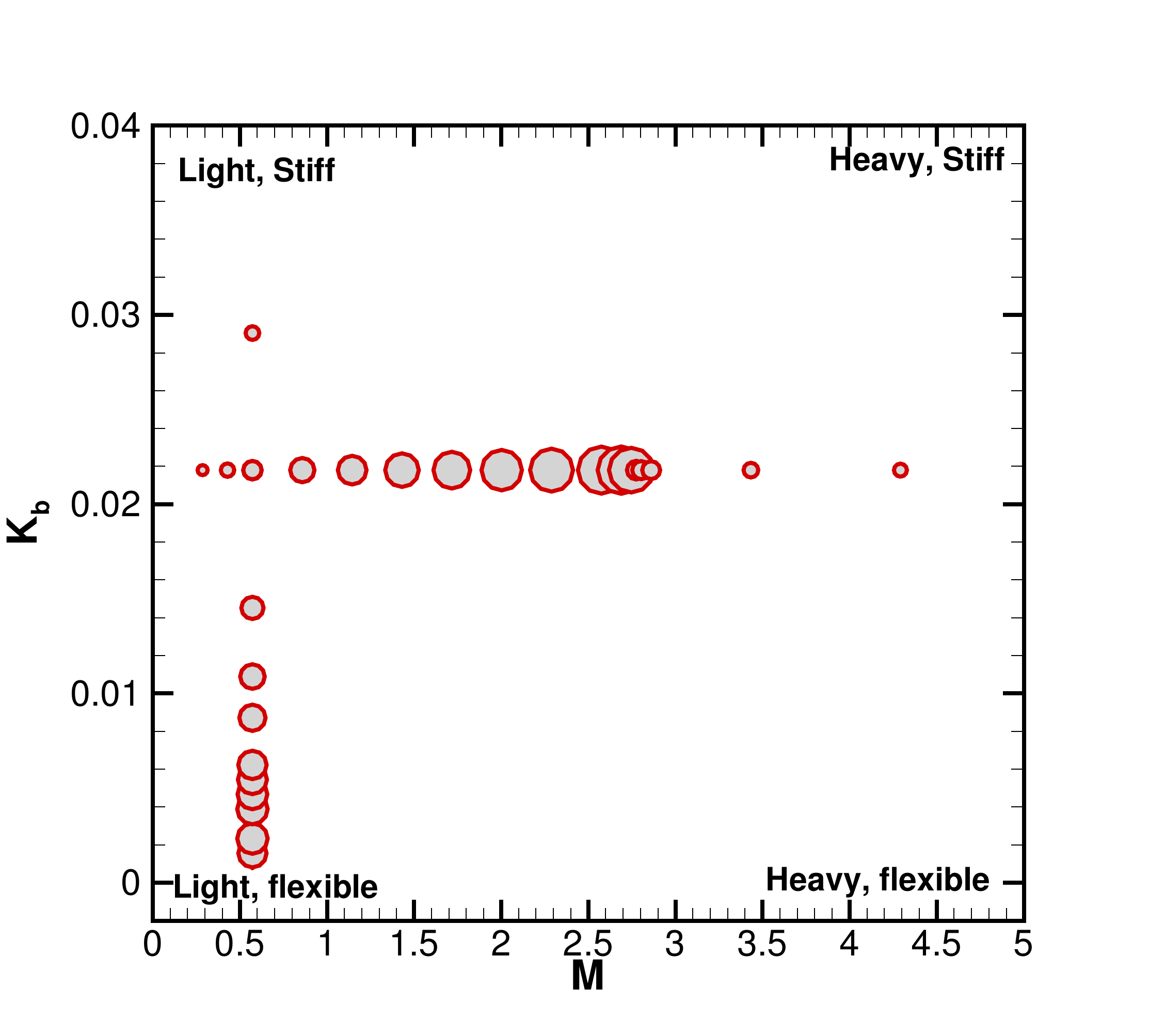}
 \caption{\rb{Qualitative comparison of the amplitude of the plate tip on $M$-$K_b$ plane for all cases. The radius of the circles are scaled with the magnitude of the computed amplitude. A large (small) $M$ and $K_b$ represent heavy (light) and stiff (flexible) plate, respectively.}}
 \label{fig:App}
\end{figure}

\rb{The present numerical data is envisioned to design MEMS-based piezoelectric energy harvesters, that harnesses available  ambient wind energy. Such miniaturized harvesters are potential candidates to replace traditional chemical batteries. In particular, they could be useful in sensors mounted on a tall bridge on which ambient fluid energy is easily available and replacement of the batteries in the sensors is expensive \cite{park2010long}. In general, $Re$ is on the order of $O$(100)-$O$(1000) in such miniaturized devices. Table 8 summarizes recent few studies which examined the design and performance of such devices. In the present study, $Re$ based on the plate length is 350, which is on the order of $Re$ used in the previous studies \cite{liu2012development, he2013micromachined, lee2019vortex}. Fig.~\ref{fig:App} plots amplitude of the plate tip on $M$-$K_b$ plane for all cases considered in the present work and the radius of the circle represents the magnitude of the amplitude. Therefore, ($M$, $K_b$) = (2.69, 0.0218) is an optimum point in Fig.~\ref{fig:App}, at which FID of the plate is maximum, implying that large energy could be harnessed at this design point. }

\section{Conclusions}\label{sec:conclusion}

We have presented the development of fluid-structure interaction (FSI) solver to simulate large-scale flow-induced dynamics of a thin elastic structure. The FSI computational approach combines a sharp-interface immersed boundary (IB) method-based flow solver and an open-source finite-element-based structure solver. The implicit coupling between the flow and structural dynamics solver has been improved with a dynamic under-relaxation scheme. The revised coupling is around two to three times faster and numerically stable, as compared to the one that uses a constant under-relaxation parameter. \rb{To bring more numerical stability to the coupling, we have implemented an additional sub-scheme in which the solver starts using constant under-relaxation for few sub-iterations if the dynamic under-relaxation scheme diverges.} The solver was validated against two-dimensional FSI benchmark problems in which a thin elastic plate is attached to a circular and square cylinder, and attains self-sustained oscillation. The present study reports new numerical data-sets while extending the FSI benchmark proposed by Turek and Hron \cite{turek2006proposal}, which could serve as additional benchmark data for future studies. In particular, we report a case with a larger mass ratio for FIV of a splitter plate in which tip-displacement is twice larger than in the FSI benchmark proposed by Turek and Hron \cite{turek2006proposal}.

We have employed the FSI solver to simulate and analyze the dynamics of an elastic splitter plate attached to a rigid circular cylinder that is subjected to two-dimensional laminar flow. The effect of mass ratio $(M)$ and bending stiffness $(K_b)$ on the FSI response are studied at $Re$ = 100. Here, $Re$ is based on free-stream velocity and cylinder diameter. We vary $M$, $K_{b}$ and $U_{R}$ in the ranges [0.143, 20], [0.0008, 0.0894] and [2.562, 30.3], respectively, noting that these ranges cover a high amplitude FSI response. The plate amplitude and oscillation frequency are found to be a function of $M$ and $K_b$. The time-varying displacement of the tip of the plate, power spectra of the displacement signal, phase plots of the plate tip displacement and the wake structure are examined to quantify the results. The largest amplitude of the plate is found to be for the lock-in region at which the natural frequency of the plate in a given fluid synchronizes with the oscillation frequency of the plate. This behavior is consistent with the classical vortex-induced vibration of a rigid cylinder. In closure, the present results provide fundamental insights into the flapping of an elastic splitter plate attached to a rigid circular cylinder, which could prove useful to the design of elastic plates for energy harvesting and thermal augmentation applications.

\section{Acknowledgements}

R.B. gratefully acknowledges financial support from the Naval Research Board (NRB), New Delhi, India through grant NRB-403/-HYD/17-18. A.K. was a recipient of the Prime Minister Fellowship Scheme for Doctoral Research, a public-private partnership between Science and Engineering Research Board (SERB), and Confederation of Indian Industry (CII). The author's host institute for research was IITB-Monash Research Academy and the partner company was Forbes Marshall Inc, Pune, India. We thank anonymous reviewers for their valuable suggestions.

\section{Conflict of Interest Statement}
On behalf of all authors, the corresponding author states that there is no conflict of interest.

\section*{References}
\bibliography{mybibfile.bib}
\pagebreak

\newpage

\begin{table}[h]
\caption{\rb{Definitions of major symbols used in the present paper.}}
\centering
\begin{tabular}{p{0.25\linewidth}p{0.8\linewidth}}
 & \\
Symbols & Definitions \\
$A_{Ytip}$ & Amplitude of Y-displacement of the tip of the plate [$A_{Ytip}^*/D^*$]\\
$b$ & Plate width [$ b^*/ D^*$]\\
$D^*$ & Diameter of the cylinder [m] \\
$E$ & Young\textsc{\char13}s modulus [$E^*$/$\rho_{f}^*{U_{\infty}^*}^2$]\\
$f_{ni}$ & Natural frequency of the plate in vacuum [${f_{ni}}^* D^*/{U_{\infty}^*}$] (eq. 5) \\
$f_{p}$ & Oscillation frequency of the plate [${f_{p}}^* D^*/{U_{\infty}^*}$] \\
$h$& Plate thickness [$ h^*/ D^*$]\\
$I^*$ & Moment of inertia of the plate cross-section [$ b^*{h^*}^3/ 12$] $[m^4]$\\
$K_b$ & Bending stiffness (eq. 1)\\
$L^*$ & Plate length [m] \\
$M$ & Mass ratio (eq. 2)\\
$\rb{N}$ & \rb{Number of triangular finite-elements in plate}\\
$Re$ & Reynolds number [$\rho_{f}^* {U_{\infty}^*} D^*/ \mu^*$]\\
$\rb{\Delta t}$ & \rb{ Time step in the simulation} \\
\rb{$u_i$} & \rb{Velocity vector}\\
$U_R$ & Reduced velocity (eq. 3)\\
$U_{\infty}^*$ & Inlet velocity [m s$^{-1}$]\\
\rb{$\mathbf{V}$} & \rb{Velocity vector of fluid-structure interface}\\
\rb{$x_i$} & \rb{Spatial coordinates}\\
\rb{$\Delta x_{min}$} & \rb{Minimum grid size in $x$-direction near the plate}\\
\rb{$\mathbf{X}$} & \rb{Position vector of fluid-structure interface}\\
$\rb{X_{tip}}$ & \rb{$x$-displacement of the tip of the plate [$X_{tip}^*/D^*$]}\\
\rb{$\Delta y_{min}$} & \rb{Minimum grid size in $y$-direction near the plate}\\
$\rb{Y_{tip}}$ & \rb{$y$-displacement of the tip of the plate [$Y_{tip}^*/D^*$]}\\
Greek symbols\\
$\mu^*$ & Dynamic viscosity [Pa.s]\\
$\rho$ & Density ratio [$\rho = {\rho_{s}}/{\rho_{f}}$]\\
Subscripts \\
$f$ & Fluid\\
$p$ & Plate\\
$ni$ & natural frequency of mode $i$\\
Superscript \\
$*$ & Dimensional quantity\\
Acronyms \\
\rb{ALE} & \rb{Arbitrary Lagrangian-Eulerian}\\
\rb{FID} & \rb{Flow-induced deformation}\\
\rb{FIV} & \rb{Flow-induced vibration}\\
\rb{FSI} & \rb{Fluid-structure interaction}\\
\rb{IB}  & \rb{Immersed boundary}\\
\end{tabular}

\end{table}

\begin{table}[h]
\caption{\rb{Details of hierarchy of grids considered for carrying out grid independence study for the fluid domain. The grid sizes in $x$- and $y$- direction is kept same where the plate is expected to move ($\Delta x_{min}= \Delta y_{min}$) and are tabulated for all the cases. The ratios of grid sizes ($\zeta$) of two successive cases ($\zeta$ = $\Delta x_{min}^k/\Delta x_{min}^{k+1}$) are also listed ($\Delta x_{min}^k$ is the grid size of case $k$). Computed amplitudes of the vibration of the plate ($A_{Ytip}$) for different grid sizes are given and L$_2$ error norms are calculated for each case relative to the finest grid.}} 
\centering 
\begin{tabular}{|c c c c c c|} 
\hline 
Case & Grid points & \rb{Grid size} & \rb{$\zeta$}  & \rb{$A_{Ytip}$} & \rb{L$_2$ error norm} \\
\hline 
\rb{1} & \rb{193 $\times$ 65}  & \rb{0.050}  & \rb{1.67}  & \rb{0.741}  &  \rb{0.014}  \\ 
\rb{2} & \rb{257 $\times$ 97}  & \rb{0.030}  & \rb{1.20}  & \rb{0.743}  &  \rb{0.012} \\ 
\rb{3} & \rb{321 $\times$ 161} & \rb{0.025}  & \rb{1.25}  & \rb{0.747}  &  \rb{0.008}  \\ 
    4  & 385 $\times$ 193 &     0.020   & \rb{1.33}  & 0.748  &  \rb{0.007} \\ 
\rb{5} & \rb{481 $\times$ 193} & \rb{0.015}  & \rb{-}     & \rb{0.755}  &  -     \\ 
\hline 
\end{tabular}
\label{table:nonlin} 
\end{table}

\begin{table}[h]
\caption{\rb{Variation in the vibration amplitude ($A_{Ytip}$) for different domain sizes. Percentage differences in the amplitude relative to the largest domain size are computed for different domains considered. } } 
\centering 
\begin{tabular}{|c c c c c c c| } 
\hline\hline 
Cases & Domain size &\rb{$S_1$ }&\rb{$S_2$}&\rb{$S_3$}&$A_{Ytip}$ & Percentage difference in $A_{Ytip}$ \\ [0.5ex] 
\hline 
1 & 19.5$D$ $\times$12$D$ &  \rb{5.5$D$}  & \rb{14$D$}   & \rb{6$D$} &\rb{0.748} & \rb{-0.95\%} \\
2 & 30$D$ $\times$12$D$  &  \rb{5.5$D$}  & \rb{25.5$D$ }& \rb{6$D$} &\rb{0.748}& \rb{-0.95\%} \\ 
3 & 40$D$ $\times$12$D$  &  \rb{5.5$D$}  & \rb{34.5$D$} & \rb{6$D$} &\rb{0.747}& \rb{-0.81\%} \\
\rb{4} & \rb{50$D$ $\times$ 20$D$}  &  \rb{5.5$D$}  & \rb{45.5$D$} & \rb{10$D$} &\rb{ 0.741} & \rb{- }\\ [1ex]
\hline 
\end{tabular}
\label{table:wes} 
\end{table}

\begin{table}[h]
\caption{ \rb{Grid-size convergence study for structure domain. The numbers of triangular finite elements ($N$) are given for the different grids tested. Computed L$_2$ errors norm in the plate amplitude ($A_{Ytip}$) for different grids with respect to the finest grid examined are also listed. }}

\centering 
\begin{tabular}{|c c c c| } 
\hline 
Cases & Number of finite-elements ($N$) & \rb{$A_{Ytip}$} & \rb{L$_2$ error norm}  \\ 
\hline 
\rb{1}& \rb{736}&	\rb{0.709}&	 \rb{0.046}\\
\rb{2}&	\rb{1262}&	\rb{0.738}&	 \rb{0.017} \\
\rb{3}&	\rb{2182}&	\rb{0.748}&  \rb{0.007}\\
\rb{4}&	\rb{3082} & \rb{0.755}&  \rb{-} \\ 

\hline 
\end{tabular}
\label{table:table43} 
\end{table}

\begin{table}[h]
\caption{\rb{Comparison of computed dimensionless amplitude ($A_{Ytip}$) and frequency ($f_{p}$) of a thin splitter plate attached on a square cylinder at $Re$ = 100 with published data. The FSI benchmark was proposed by Wall and Ramm \cite{wall1998fluid}.}} 
\centering 
\begin{tabular}{|c c c| } 
\hline\hline 
Study & $f_{p}$ & $A_{Ytip}$ 
 \\ [0.5ex] 
\hline 
Present work & 0.0637 & 0.97 \\ 
Olivier et al. \cite{olivier2009fluid} & 0.0617&0.95 \\
Habchi et al. \cite{habchi2013partitioned} & 0.0634 & 1.02 \\
Wall and Ramm \cite{wall1998fluid} &0.0581& 1.22 \\ [1ex] 
\hline 
\end{tabular}
\label{table:square} 
\end{table}

\begin{table}[h]
\caption{Input parameters for the set of simulations to studying the effect of mass ratio, $M$, for fixed bending stiffness, $K_b$.} 
\centering 
\begin{tabular}{|c c c c c c c c c| } 
\hline 
Case number & $E$ & $h$ & $\rho$ & $M$& $K_{b}$ & $U_R$ &$A_{Ytip}$ & $f_p$ \\ [0.2ex] 
\hline 
1 & 1400 & 0.2 & 2.5 & 0.143 & 0.0218 & 2.562 & 0 & 0\\
2 & 1400 & 0.2 & 5.0 & 0.286 & 0.0218 & 3.623 & 0.41& 0.183 \\
3 & 1400 & 0.2 & 7.5 & 0.429 & 0.0218 & 4.437& 0.55& 0.167 \\
4 & 1400 & 0.2 & 10.0 & 0.572 & 0.0218 & 5.123& 0.75& 0.158 \\
5 & 1400 & 0.2 & 15.0 & 0.859 & 0.0218 & 6.275& 0.99& 0.137 \\
6 & 1400 & 0.2 & 20.0 & 1.145 & 0.0218 & 7.246 & 1.17& 0.122\\
7 & 1400 & 0.2 & 25.0 & 1.431 & 0.0218 & 8.101& 1.37 & 0.107\\
8 & 1400 & 0.2 & 30.0 & 1.717 & 0.0218 & 8.874& 1.5& 0.096 \\
9 & 1400 & 0.2 & 35.0 & 2.003 & 0.0218 & 9.585& 1.65& 0.085 \\
10 & 1400 & 0.2 & 40.0 & 2.289 & 0.0218 & 10.247& 1.75& 0.079\\
11 & 1400 & 0.2 & 45.0 & 2.576 & 0.0218 & 10.869& 1.92& 0.076 \\
12 & 1400 & 0.2 & 47.0 & 2.690 & 0.0218 & 11.107& 1.93& 0.0732 \\
13 & 1400 & 0.2 & 48.0 & 2.747 & 0.0218 & 11.225& 1.82& 0.073 \\
14 & 1400 & 0.2 & 48.5 & 2.775 & 0.0218 & 11.283 & 0.76& 0.077\\
15 & 1400 & 0.2 & 49.0 & 2.804 & 0.0218 & 11.341& 0.74& 0.076 \\
16 & 1400 & 0.2 & 50.0 & 2.861 & 0.0218 & 11.456& 0.73& 0.073 \\
17 & 1400 & 0.2 & 60.0 & 3.434 & 0.0218 & 12.550& 0.60& 0.070\\
18 & 1400 & 0.2 & 75.0 & 4.292 & 0.0218 & 14.031 & 0.53& 0.067\\
19 & 1400 & 0.2 & 90.0 & 5.150 & 0.0218 & 15.370 & 0.46& 0.063\\
20 & 1400 & 0.2 & 100.0 & 5.723 & 0.0218 & 16.202& 0.45& 0.061 \\
21 & 1400 & 0.2 & 125.0 & 7.153 & 0.0218 & 18.114 & 0.30& 0.058\\
22 & 1400 & 0.2 & 150.0 & 8.584 & 0.0218 & 19.843& 0.27& 0.059 \\
23 & 1400 & 0.2 & 200.0 & 11.445 & 0.0218 & 22.913& 0.24& 0.048 \\
24 & 1400 & 0.2 & 250.0 & 14.306 & 0.0218 & 25.617& 0.17& 0.045 \\
25 & 1400 & 0.2 & 300.0 & 17.167 & 0.0218 & 28.062& 0& 0\\
26 & 1400 & 0.2 & 350.0 & 20.029 & 0.0218 & 30.311& 0& 0\\[1ex]
\hline 
\end{tabular}
\label{table:Mass} 
\end{table}

\begin{table}[h]
\caption{Input parameters for the set of simulations to studying the effect of bending stiffness, $K_b$, for fixed mass ratio, $M$.} 
\centering 
\begin{tabular}{|c c c c c c c c c| } 
\hline 
Case number & $E$ & $h$ & $\rho$ & $M$& $K_{b}$ & $U_R$&$A_{Ytip}$ & $f_p$ \\ [0.2ex] 
\hline 
27 & 50 & 0.2 & 10.0 & 0.572 & 0.0008 & 27.111 & 0.010 & 0.116 \\
28 & 100 & 0.2 & 10.0 & 0.572 & 0.0016 & 19.170 & 1.150 & 0.087 \\
29 & 150 & 0.2 & 10.0 & 0.572 & 0.0023 & 15.652 & 1.242 & 0.098 \\
30 & 250 & 0.2 & 10.0 & 0.572 & 0.0039 & 12.124 & 1.244 & 0.095 \\
31 & 300 & 0.2 & 10.0 & 0.572 & 0.0047 & 11.068 & 1.214 & 0.102 \\
32 & 350 & 0.2 & 10.0 & 0.572 & 0.0055 & 10.247 & 1.185 & 0.104 \\
33 & 400 & 0.2 & 10.0 & 0.572 & 0.0062 & 9.585 & 1.159 & 0.110 \\
34 & 560 & 0.2 & 10.0 & 0.572 & 0.0087 & 8.101 & 1.070 & 0.121 \\
35 & 700 & 0.2 & 10.0 & 0.572 & 0.0109 & 7.246 & 1.000 & 0.122 \\
36 & 933 & 0.2 & 10.0 & 0.572 & 0.0145 & 6.276 & 0.905 & 0.137 \\
4 & 1400 & 0.2 & 10.0 & 0.572 & 0.0218 & 5.123 & 0.753 & 0.158 \\
37 & 1866 & 0.2 & 10.0 & 0.572 & 0.0291 & 4.438 & 0.560 & 0.173 \\
38 & 2800 & 0.2 & 10.0 & 0.572 & 0.0436 & 3.623 & 0.000 & 0.000 \\
\hline 
\end{tabular}
\label{table:kb} 
\end{table}

\begin{table}[h]
\caption{Recent previous studies which examined wind energy based miniaturized energy harvesters. The range of flow velocity and $Re$ based on characteristic length in the study are listed in the table. Kinematic viscosity of air is taken as 1.54 $\times$ 10$^{-5}$ m$^2$ s$^{-1}$ in ambient conditions for calculations of $Re$. In the present study, $Re$ based on the plate length is 350.}
\centering
\scriptsize
\begin{tabular}{|c c c c c| } 
\hline 
\rb{Study by} & \rb{Mode of} & \rb{Cantilever length/} & \rb{Wind speed (m s$^{-1}$)} & \rb{$Re$ based on $l^*$} \\
 & \rb{energy harvesting} & \rb{cylinder diameter ($l^*$, mm) }& &  \\ 
\hline 
\rb{Liu et al. \cite{liu2012development}} & \rb{FIV of a piezoelectic cantilever} & \rb{3.0} & \rb{3.9-15.6} & \rb{760-3069} \\
\rb{He et al. \cite{he2013micromachined}} & \rb{FIV of a piezoelectic cantilever} & \rb{3.0} & \rb{0.6-16.3}    & \rb{1137-3175}\\ 
 & \rb{with proof mass} &  &   & \\ 

\rb{Lee et al. \cite{lee2019vortex}}& \rb{VIV of a cylinder }& \rb{2.0} & \rb{1-6 } & \rb{130-779 }\\
\hline 
\end{tabular}
\label{table:time} 
\end{table}

\clearpage

\clearpage

\clearpage

\clearpage

\clearpage

\clearpage

\clearpage

\clearpage

\clearpage

\clearpage

\end{document}